\documentclass[12pt]{article}

\usepackage{scicite}

\usepackage{times}

\newenvironment{sciabstract}{
\begin{quote} \bf}
{\end{quote}}

\usepackage{graphicx}
\usepackage{siunitx}

\usepackage{listings}
\usepackage{xcolor}
\definecolor{codegreen}{rgb}{0,0.6,0}
\definecolor{codegray}{rgb}{0.5,0.5,0.5}
\definecolor{codepurple}{rgb}{0.58,0,0.82}
\definecolor{backcolour}{rgb}{0.95,0.95,0.92}
\lstdefinestyle{mystyle}{
    backgroundcolor=\color{backcolour},   
    commentstyle=\color{codegreen},
    keywordstyle=\color{magenta},
    numberstyle=\tiny\color{codegray},
    stringstyle=\color{codepurple},
    basicstyle=\ttfamily\footnotesize,
    breakatwhitespace=false,         
    breaklines=true,                 
    captionpos=b,                    
    keepspaces=true,                 
    numbers=left,                    
    numbersep=5pt,                  
    showspaces=false,                
    showstringspaces=false,
    showtabs=false,                  
    tabsize=2
}
\usepackage{physunits}

\newcommand{\apos}[1]{`#1'}

\usepackage{amsthm}

\usepackage{amsmath}
\usepackage{comment}
\usepackage{soul}

\DeclareSIUnit{\SLPM}{SLPM}
\DeclareSIUnit{\BL}{BL}
\DeclareSIUnit{\GA}{GA}

\title{Embodying mechano-fluidic memory in soft machines to program behaviors upon interactions}

\author
{Alberto Comoretto$^{1}$, Tanaya Mandke,$^{1}$ Johannes T.B. Overvelde$^{1,2\ast}$\\
\\
\normalsize{$^{1}$Autonomous Matter Department, AMOLF,}\\
\normalsize{Science Park 104, 1098 XG Amsterdam, The Netherlands}\\
\normalsize{$^{2}$Institute for Complex Molecular Systems,}\\
\normalsize{Department of Mechanical Engineering, Eindhoven University of Technology,}\\
\normalsize{PO Box 513, 5600 MB Eindhoven, The Netherlands}\\
\\
\normalsize{$^\ast$Correspondence:  overvelde@amolf.nl.}
}

\date{}

\begin{document} 

\baselineskip24pt

\maketitle 

\begin{sciabstract}
Soft machines display shape adaptation to external circumstances due to their intrinsic compliance. To achieve increasingly more responsive behaviors upon interactions without relying on centralized computation, embodying memory directly in the machines' structure is crucial. Here, we harness the bistability of elastic shells to alter the fluidic properties of an enclosed cavity, thereby switching between stable frequency states of a locomoting self-oscillating machine. To program these memory states upon interactions, we develop fluidic circuits surrounding the bistable shell, with soft tubes that kink and unkink when externally touched. We implement circuits for both long-term and short-term memory in a soft machine that switches behaviors in response to a human user and that autonomously changes direction after detecting a wall. By harnessing only geometry and elasticity, embodying memory allows physical structures without a central brain to exhibit autonomous feats that are typically reserved for computer-based robotic systems. 
\end{sciabstract}

\section*{Keywords}

Soft machines $|$ Mechanical memory $|$ Mechanical instabilities $|$ Elastic shells $|$ Fluidic circuits

\section*{Introduction}

Soft robots are being developed for autonomous operation in the complex real world \cite{design_fabrication_soft}. Often inspired by biological systems \cite{bioinspired_softrobots}, soft robots passively adapt to external stimuli due to their intrinsic compliance \cite{whitesides_softrobotics}. This passive adaptability at the level of material and structure, often called mechanical or embodied intelligence \cite{how_the_body}, enables soft robots to accomplish tasks such as grasping a wide variety of objects using the same gripper \cite{jamming_gripper}, walking over uneven terrain \cite{legged_robot_unstructured}, resisting from external damage \cite{untethered_multigait, 3D_printed_tauber}, and even self-healing \cite{self_heal}.

Their natural counterpart, animals, achieve feats that are more complex than mechanical adaptation: they often dynamically change their behaviors in response to external stimuli. For instance, sea stars typically slowly explore the environment in search of food, but when threatened by predators, they suddenly enter a fast galloping gait as an escape response \cite{seastar_heydari, cooperative_seastar}. Salamanders switch between two stable locomoting gaits: undulatory swimming in the water and slower stepping motions on the ground \cite{salamander_animal, salamander_robot}. Even the relatively simple organism \emph{Caenorhabditis elegans} switches between basic behavioral states of locomotion, such as forward moving and turning, depending on the surroundings, previous experiences, and internal factors \cite{behavioral_states}. In general, this switching of behaviors can be seen as a form of memory \cite{mechanical_memory}, where each adopted behavior is a stable memory state.

In an effort to provide soft machines with this kind of responsiveness, researchers developed artificial systems that harness structural phenomena to passively exhibit distinct behaviors depending on external cues. Mechanical robots, in which geometric nonlinearities combined with elasticity lead to reprogrammable mechanisms with multi-welled energy landscapes, continuously change the internal activation sequences as a direct response to interactions \cite{reprogrammablesequencing}. Soft twisting liquid crystal elastomers \cite{twisting_LCE} and elasto-active structures \cite{bucklebots} exploit environmental interactions and passive shape reconfiguration to solve mazes. Soft modular machines sense and respond to different external stimuli by harnessing responsive materials \cite{modular_strategy_embodied_control}. In previous work, we introduced soft machines with self-oscillating limbs \cite{physical_synchronization} that passively tune their synchronization pattern depending on external cues through implicit coupling with the surrounding medium. These exciting initial advances point toward embodying switchable behaviors within the physical structure of the machine itself. In this direction, there is ample room for further investigation in understanding and then utilizing structural phenomena not only for mechanical shape adaptation but also for behavioral adaptation. In particular, memory effects to selectively program desired, stable behaviors upon external interaction remain largely unexplored.

Focusing in this work on soft fluidic machines, fluidic circuits \cite{hardware_methods_fluidic_control}, consisting of interconnections of pneumatic tubings \cite{harnessing_viscous_flow, nonuniform_pressure_robot, kink_valves} and nonlinear inflatable elements \cite{amplifying_with_instabilities} and valves \cite{cheap_valves, soft_bistable_valve, monolithic_soft_fluidic, luuk, mousa_frequency_control, 3D_printed_tauber}, represent a promising platform for implementing behaviors at the centimeter scale. In fact, multiple tools have been developed and are available to designers, including viscous flow \cite{harnessing_viscous_flow, nonuniform_pressure_robot}, snap-through instabilities \cite{amplifying_with_instabilities, gorissen_sequencing, piano_gorissen}, transistors \cite{digital_logic_soft} and oscillators \cite{soft_bistable_valve, luuk, bro, coexistence, mousa_frequency_control, physical_synchronization}. Fluidic circuits have proven effective for a variety of behaviors, including sequential activation of soft fingers \cite{luuk}, automatic gripping \cite{soft_bistable_valve, monolithic_soft_fluidic} and open-loop gait control for walking robots \cite{luuk, 3D_printed_tauber, gorissen_sequencing, drotman_turtle, bro}. Despite the progress at the component and circuitry level, integration in autonomous systems capable of responding and adapting to changing environmental circumstances remains elusive. One exception consists in changing behavior in response to external cues only once, as seen in a walker that changes locomotion direction when obstructed \cite{drotman_turtle}, and in an extensible gripper that we developed that transitions from searching to retrieving a sensed object \cite{pneumatic_coding}.

Here, by instilling memory effects in the physical body of soft fluidic machines, we enable the programmability of stable behaviors upon repeated interactions with the surroundings. We start from a soft robotic crawler, to which we provide memory by harnessing the bistability of elastic shells \cite{Taffetani2018}. Through fluidic circuits that surround the shells, we show both long-term and short-term memory of touch interactions. After each interaction, the memory state is rewritten, and the response of the machine changes accordingly. Equipped with fluidic antennae, the machine detects the presence of obstacles, memorizes this information, and responds by autonomously steering away. By introducing such a physical form of memory, we expand the repertoire of design tools for autonomous soft machines \cite{reprogrammablesequencing, twisting_LCE, bucklebots, modular_strategy_embodied_control, physical_synchronization}, which can now remember interactions after they occur, in addition to passively responding to them.

\section*{Results}

\paragraph*{A self-oscillating soft fluidic machine}

\begin{figure}[t!]
\centering
\includegraphics[width=16.5cm]{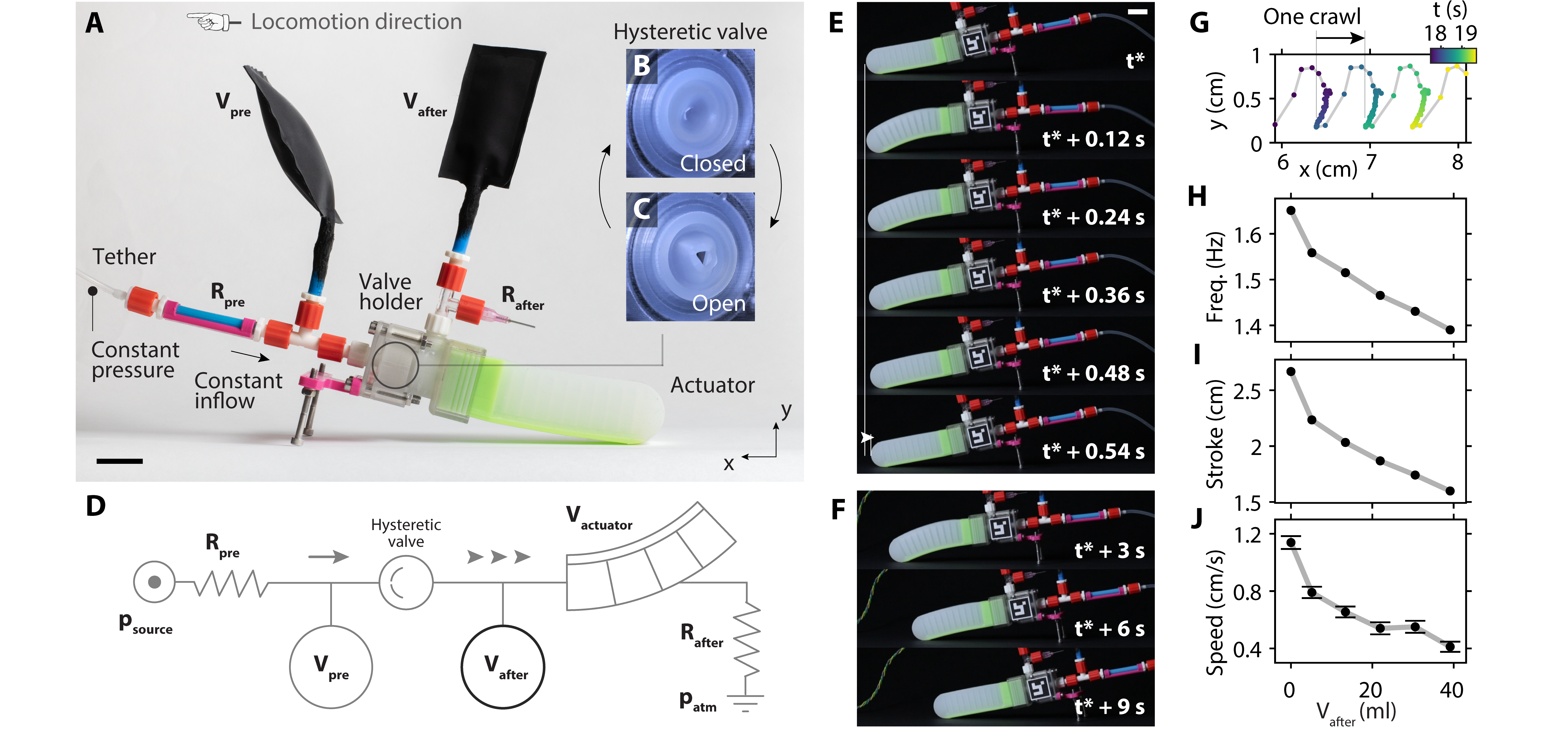}
\par\medskip
    \caption{\label{fig1} \textbf{A self-oscillating soft fluidic machine.} (\textbf{A}) A soft crawling machine is equipped with a bending actuator, pre- and after-capacitances ($V_\textrm{pre}$ and $V_\textrm{after}$), pre- and after-resistances ($R_\textrm{pre}$ and $R_\textrm{after}$), and a hysteretic valve mounted in a rigid holder. A single tether provides constant pressure to the pre-resistance. Given the approximately constant inflow from the pre-resistance, the soft hysteretic valve oscillates between (\textbf{B}) closed and (\textbf{C}) open states, enabling pulsatile actuation of the bending actuator. (\textbf{D}) Schematic of the fluidic circuit embedded in the machine. (\textbf{E}) Locomotion of the crawler during one oscillation cycle lasting $\SI{0.54}{\second}$.  (\textbf{F}) Location of the crawler after $\SI{3}{\second}$, $\SI{6}{\second}$, and $\SI{9}{\second}$. (\textbf{G}) Horizontal and vertical coordinates of the rigid holder of the locomoting machine. (\textbf{H}) Influence of the after-capacitance geometrical volume $V_{\textnormal{after}}$ on the oscillation frequency of the actuator, (\textbf{I}) vertical stroke of the actuator, and (\textbf{J}) speed of the machine. All scale bars are \SI{2}{\centi\meter}. See also Fig.~\ref{fig_hysteretic_valve} and Fig.~\ref{fig_varyfluidicparameters}. 
    }
\end{figure}

We start by designing a self-oscillating locomoting machine consisting of a single pneumatic bending actuator (Fig.~\ref{fig1}A). Given a constant pressure source, the machine crawls in a pulsatile fashion, because it is cyclically activated through a hysteretic valve, mounted inside the machine, previously developed in our group \cite{luuk, coexistence}. Given constant inflow, the valve oscillates between a closed state (Fig.~\ref{fig1}B) and an open state (Fig.~\ref{fig1}C). While oscillating, the valve goes through several stages (Fig.~\ref{fig_hysteretic_valve}). \emph{i)} When closed, no air flow is delivered to the actuator, and pressure upstream increases. \emph{ii)} When a critical pressure is reached, the valve snaps to the open state, allowing flow to the actuator, decreasing pressure upstream. \emph{iii)} When a low critical pressure difference is reached, the valve snaps back to the closed state. This hysteresis in opening and closing results in cyclic inflation and deflation of the actuator, leading to forward locomotion.

The fluidic circuit carried by, and activating, the machine (Fig.~\ref{fig1}D) is characterized by six key physical parameters. \emph{i)} The pressure source value together with the \emph{ii)} pre-resistor with resistance $R_\textrm{pre}$ determine the amount of inflow to the valve. That is, both a larger pressure source and a smaller pre-resistance lead to higher inflow. \emph{iii)} The pre-capacitance with volume $V_\textrm{pre}$ is responsible for the timescale of charging, where larger volume leads to longer charging. After the valve, we place  \emph{iv)} the actuator with geometric volume $V_\textrm{actuator}$, \emph{v)} the after-capacitance $V_\textrm{after}$, and  \emph{vi)} the after-resistance $R_\textrm{after}$, responsible for the amount of inflation of the actuator and the discharge time \cite{luuk}.

Practically, we build the pre-resistor as a custom-made silicone tube with a small inner diameter of \SI{0.35}{\milli\meter}, outer diameter \SI{5}{\milli\meter} and length \SI{40}{\milli\meter} through injection molding; the after-resistor is an off-the-shelf needle ($20$ Gauge, \SI{0.5}{\inch} length); the capacitors are thin inextensible pouches custom-made through heat-sealing of nylon-coated thermoplastic polyurethane (TPU) sheets \cite{arfaee_pouch}; the pneumatic bending actuator is a two-material PneuNet custom-made through injection-molding \cite{shibo} (Methods).

The pulsatile inflation of the actuator induced by the fluidic circuit causes the machine to move forward at a speed of \SI{0.76}{\centi\meter\per\second} ($2.85$ body lengths per minute) (Fig.~\ref{fig1}E,F) through consecutive crawling motion of $\sim\SI{5.3}{\milli\meter}$ each ($0.03$ body lengths for each cycle) (Fig.~\ref{fig1}G). This single-input crawling machine is purposefully simple in its behavior, that is, moving forward at a specific average speed. Throughout the article, we will use this unit and its associated fluidic circuit as a building block to construct more complex behaviors by adding memory and sensing features to the circuit and assembling multiple units together.

As a start, we observe that varying the physical parameters of the circuit directly results in a change of behavior of the machine. For example, in a benchtop experiment with the actuator not interacting with the ground (Methods), increasing the volume of the after-capacitance from \SI{0}{\milli\liter} to \SI{40}{\milli\liter} results in the activation frequency decreasing from \SI{1.65}{\hertz} to \SI{1.39}{\hertz} (Fig.~\ref{fig1}H) and the vertical stroke of the actuator lowering from \SI{2.7}{\centi\meter} to \SI{1.6}{\centi\meter} (Fig.~\ref{fig1}I). This change in after-capacitance volume has a direct consequence on the speed of the locomoting machine. Adding an after-capacitance with volume $\SI{40}{\milli\liter}$ causes the machine to crawl slower at \SI{0.4}{\centi\meter\per\second}, at approximately one third of the original speed (\SI{1.1}{\centi\meter\per\second}) (Fig.~\ref{fig1}J). Varying other physical parameters also affect the behavior. For instance, increasing the pre-capacitance from \SI{0}{\milli\liter} to \SI{70}{\milli\liter} leads to a higher vertical stroke of the actuator (Fig.~\ref{fig_varyfluidicparameters}).

\paragraph*{Embodying memory via a bistable mechano-fluidic capacitor}

To introduce memory in the system, we provide bistability to a physical parameter of the self-oscillating circuit, so that the crawling behavior also becomes bistable. In our specific case, we focus on the after-capacitance, as the behavior of the machine is particularly sensitive to this parameter. This is because increasing the after-capacitance leads to a decrease in both the actuator's stroke and frequency, while increasing the pre-capacitance leads to an increase in stroke but a decrease in frequency, with a negligible net change in speed (Fig.~\ref{fig_varyfluidicparameters}).

\begin{figure}[t!]
\centering
\includegraphics[width=16.5cm]{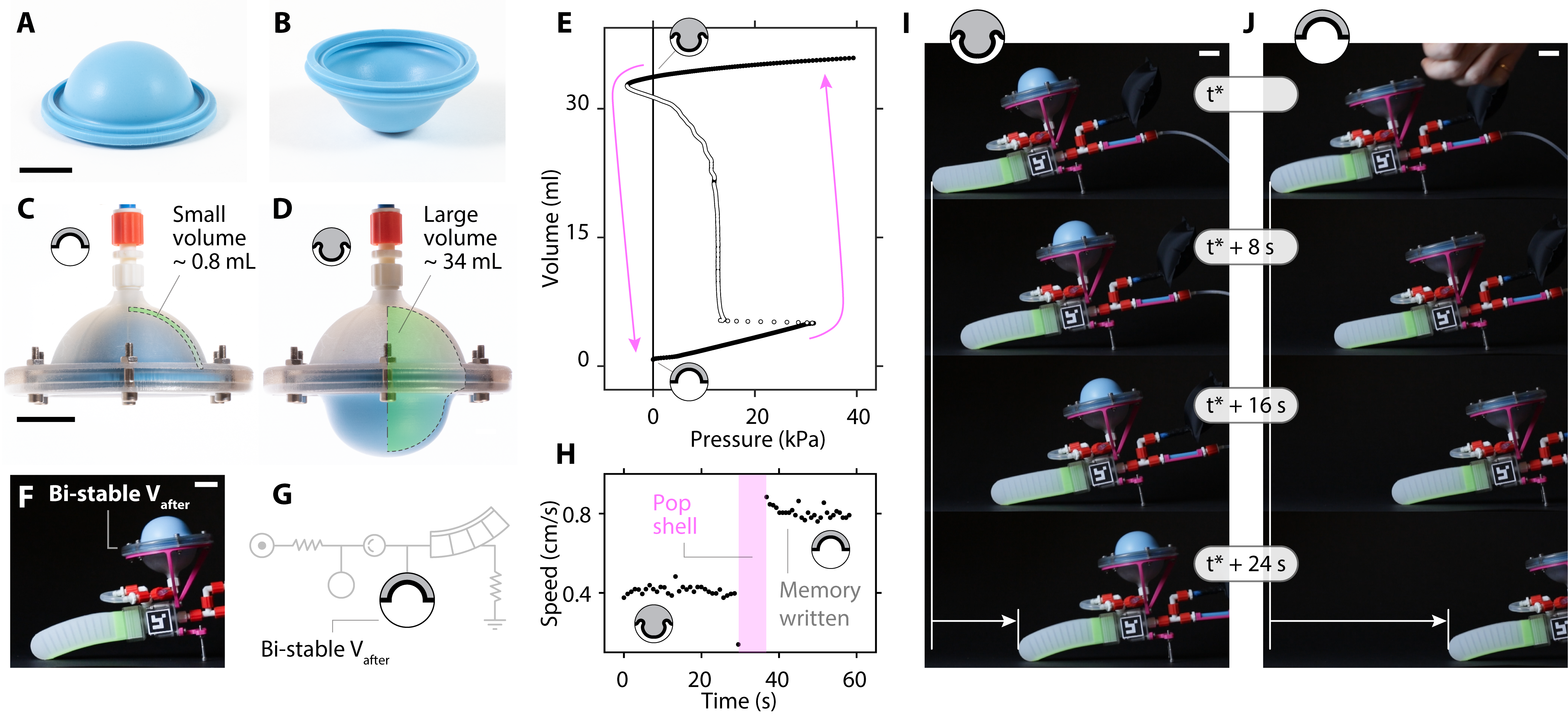}
\par\medskip
    \caption{\label{fig2} \textbf{Embodying memory via a bistable mechano-fluidic capacitor} An elastomeric shell with thickness \SI{3}{\milli\meter}, base width \SI{55}{\milli\meter}, and shallowness angle \SI{80}{\degree} displays two stable states, (\textbf{A}) a rest state and (\textbf{B}) a popped state. When the soft shell is mounted on a rigid shell-shaped holder, the geometric volume between the two shells also displays two states, (\textbf{C}) a small volume state and (\textbf{D}) a large volume state. (\textbf{E}) Increasing (or decreasing) pressure above (or below) the critical snap-through (or snap-back) pressure allows for switching between the two states. (\textbf{F}) We physically mount the bistable capacitor on top of the crawler, (\textbf{G}) fluidically connected as an after-capacitance. (\textbf{H}) We write the memory state by popping the shell while the machine crawls so that the machine switches from the first stable state with speed $\sim\SI{0.4}{\centi\meter\per\second}$ to the second stable state with speed $\sim\SI{0.8}{\centi\meter\per\second}$. Snapshots of the crawling machine in (\textbf{I}) the low-speed state and in (\textbf{J}) the high-speed state. All scale bars are \SI{2}{\centi\meter}. See also Fig.~\ref{fig_dome_design} and Video~S1.}
\end{figure}

To provide bistability to the after-capacitance, we use elastic shells, as they are well-studied structures and exhibit rich nonlinear behavior \cite{Liu2022,Pandey2014,Taffetani2018,gorissen_jumper,soft_bistable_valve,Abbasi2021,Lee2016,bistable_shells}. The design parameters such as thickness, base width, and shallowness angle (Fig.~\ref{fig_dome_design}) can be tuned \cite{soft_bistable_valve} so that the shell displays bistability \cite{Taffetani2018}, with a ‘rest' stable state (Fig.~\ref{fig2}A) and a ‘snapped' stable state (Fig.~\ref{fig2}B). We mount the soft shell in a rigid shell-shaped holder, obtaining a fluidic capacitor with a relatively small (compared to the soft actuator) geometric volume of $\sim\SI{0.8}{\milli\liter}$ enclosed between the two shells when the elastic shell is in its rest state (Fig.~\ref{fig2}C). The second stable state of the bistable capacitor is accessed by popping the shell to the snapped state, obtaining a relatively large geometric volume ($\sim\SI{34}{\milli\liter}$) (Fig.~\ref{fig2}D).

The behavior of the bistable capacitor upon inflation is highly nonlinear. The pressure-volume curve of the capacitor (Fig.~\ref{fig2}E) highlights two stable regimes that cross the zero-pressure line, which is essential to enable bistability. Increasing (or decreasing) pressure past the critical snap-through (or snap-back) pressure allows for switching from one stable regime to the other (Fig.~\ref{fig2}E, pink arrows). Once the system is placed in one regime, removing pressure results in the capacitor being stable at one of the two states ($\SI{0.8}{\milli\liter}$ or $\SI{34}{\milli\liter}$). Note how this highly nonlinear, non-monotonic behavior fundamentally differs from capacitors (i.e., actuators) typically used in fluidic circuits \cite{luuk}. Such capacitors are characterized by a monotonic pressure-volume curve, even when displaying nonlinear behaviors, such as stiffening or softening (Fig~\ref{fig_pVcurves}).

We physically mount the bistable capacitor on the crawler (Fig.~\ref{fig2}F), fluidically connected in the circuit as after-capacitance $V_\textrm{after}$ (Fig.~\ref{fig2}G). We start by setting the bistable capacitor to the snapped state with large volume (Fig.~\ref{fig2}F). When we provide a constant pressure of \SI{1.3}{\bar} to the tether, the machine locomotes at a speed of $\sim\SI{0.4}{\centi\meter\per\second}$ (Fig.~\ref{fig2}H,I). After $\sim\SI{30}{\second}$, we manually pop the shell to the rest state, so that the capacitance snaps to the small-volume state (Video~1). After this external interaction, the machine is moving at a higher speed of $\sim\SI{0.8}{\centi\meter\per\second}$ (Fig.~\ref{fig2}H,J), as expected from Figure~\ref{fig1}, because of the lower after-capacitance. Crucially, the bistability of the fluidic after-capacitance directly results in two stable outcome behaviors of the machine, wich we refer to as memory, as the state of the system reflects past interactions.

\paragraph*{Fluidic circuits for writable long-term and short-term memory}

\begin{figure}[t!]
\centering
\includegraphics[width=16.5cm]{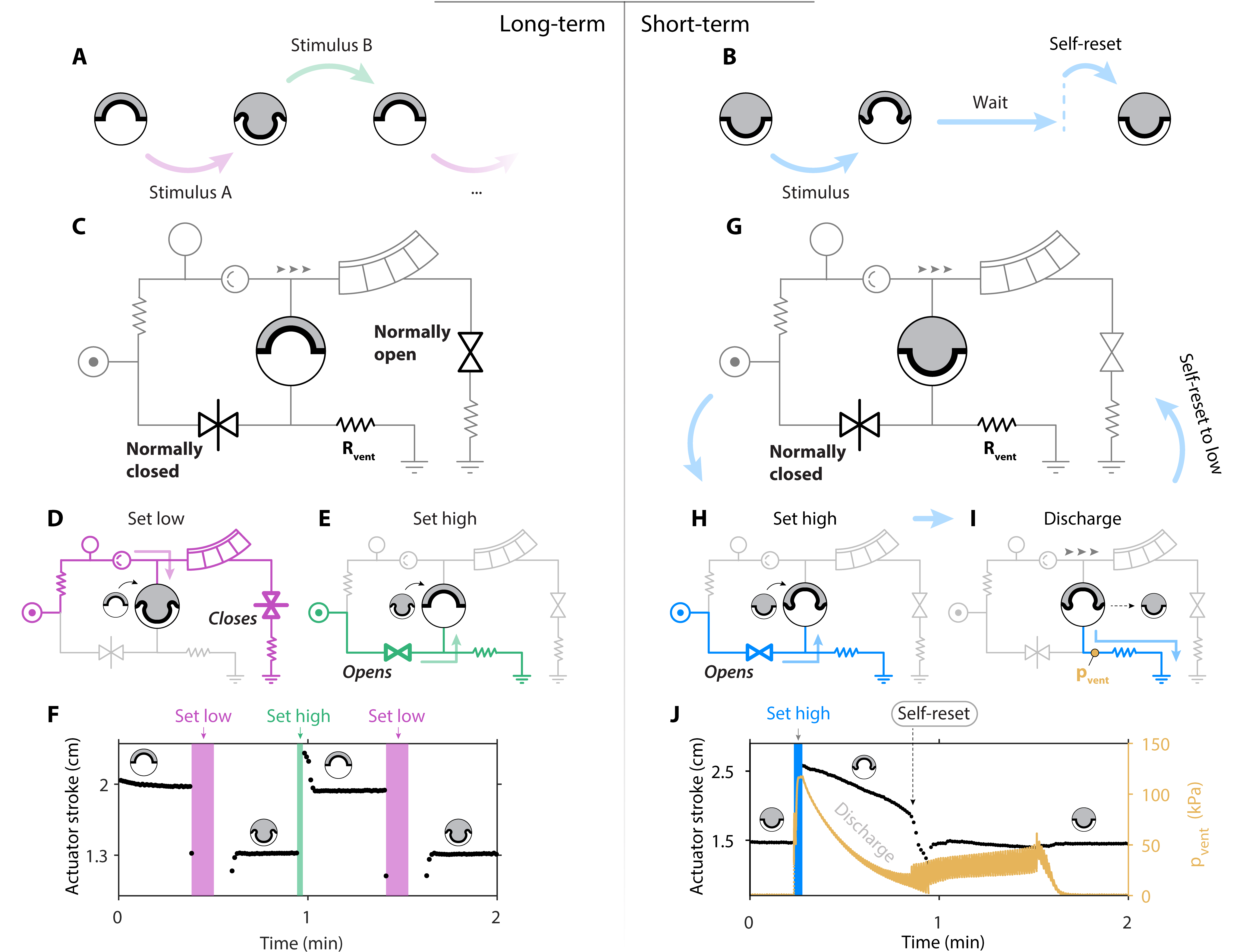}
\par\medskip
    \caption{\label{fig3} \textbf{Fluidic circuits for writable long-term and short-term memory.} (\textbf{A}) Long-term memory involves programming stable states given occurring stimuli. (\textbf{B}) Short-term memory stores information of the past stimulus for a determined amount of time, and then self-resets to the initial state. (\textbf{C}) The long-term memory circuit is equipped with a normally open (NO) valve, a normally closed (NC) valve, and a venting resistance $R_{\textrm{vent}}$ placed at the bottom chamber line. (\textbf{D}) We set the system to the state with high after-capacitance by closing the NO valve, thereby pressurizing the top chamber and snapping the shell. (\textbf{E}) We set the system to the other state by opening the NC valve, resulting in snap-back of the shell. (\textbf{F}) Actuator stroke in time, with consecutive events of writing memory. (\textbf{G}) The short-term memory circuit is equivalent to the long-term memory circuit, with the only differences being the orientation of the shell in its rest state and the NO valve not being used. (\textbf{H}) We write the memory by opening the NC valve, resulting in the shell snapping. (\textbf{I}) After this writing action, the bottom chamber is pressurized ($p_{\textnormal{vent}}$), and air leaks through the venting resistance, causing a pressure discharge in time, until the shell snaps back, causing a self-reset of the system to the low-stroke state in \textbf{G}. (\textbf{J}) Actuator stroke (black) and $p_{\textnormal{vent}}$ (yellow) in time. See also Fig.~\ref{fig_shortterm_details}, Fig.~\ref{fig_varyshortterm}, and Video~S2.}
\end{figure}

We demonstrated a bistable behavior through a mechano-fluidic memory element. So far, the memory state could be written only once by manually popping the shell from the snapped state to the rest state. In addition, we needed a relatively high external force to write the memory state ($\sim\SI{13}{\newton}$, given the surface area of the human thumb $\sim\SI{3.2}{\centi\meter\squared}$ \cite{thumb} and the internal pressure $\sim\SI{40}{\kilo\pascal})$. We now aim to repeatably change behavior upon consecutive limited-power interactions with the environment. We develop fluidic circuits around the bistable capacitor to obtain long-term memory (Fig.~\ref{fig3}A) and short-term memory  (Fig.~\ref{fig3}B) that require lower power to switch states. Note that we refer to the concepts of short- and long-term memory from a behavioral neuroscience point of view: short-term memories of the stimuli are temporary and last for a short amount of time before they fade out, while long-term memory refers to permanently lasting memories of events \cite{shortterm_memory, psychology_book}.

We develop a long-term memory circuit (Fig.~\ref{fig3}C) by building upon the circuit with the bistable capacitor in Figure~\ref{fig2}G. \emph{i)} We add a branch in parallel to the source, connected to the bottom chamber of the memory element, opposite to the actuator. \emph{ii)} In this new branch, we introduce a normally closed (NC) valve upstream and a venting resistance $R_\textrm{vent}$ downstream ($22$ Gauge, \SI{0.5}{\inch}). \emph{iii)} We add a normally open (NO) valve in the actuator branch, between the actuator and the downstream resistance. We can interact with this circuit by closing (and opening) the NO (and NC) valves. To design and test these circuits, we make use of solenoid valves controlled with an input/output electronic board (Methods), which we will replace in later sections with soft tubes that kink and unkink \cite{kink_valves}.

By closing the NO valve, the actuator line pressurizes (Fig.~\ref{fig3}D). When the pressure reaches the snap-through pressure of the shell, the shell snaps to the other state, and the capacitor is in the large volume state. This, in turn, causes the system behavior to operate at low frequency and stroke. Then, we can open the NC valve (Fig.~\ref{fig3}E), so that the bottom line pressurizes. When the pressure difference between the bottom and top lines reaches the snap-back pressure of the shell, the shell snaps back to the rest state, causing the system behavior to operate at high frequency and stroke. Each time the memory is set to either state, the reached state is stable until the next interaction, as a consequence of the bistability of the shell (Fig.~\ref{fig3}F and Video~2). Therefore, each time a stimulus occurs (opening/closing of NC/NO valves), the system expresses the memory of the past stimulus through its current behavior (frequency and stroke).

Starting from the design of the long-term memory circuit, we develop a short-term memory circuit (Fig.~\ref{fig3}G). This circuit is equivalent to the long-term memory circuit, with the only differences being a higher venting resistance ($32$ Gauge, \SI{0.25}{\inch} needle), and the shell flipped so that, in the rest state, the chamber connected to the actuator line is in the high-capacitance state (Fig.~\ref{fig3}G, grey area above the shell). By opening the NC valve, the pressure in the bottom line increases until the shell snaps, and the system behavior is set to the state with high frequency and stroke (Fig.~\ref{fig3}H).

After this initial interaction that sets the memory element to the other state, a seemingly counterintuitive behavior occurs, resulting in short-term memory. Given the bistability of the shell, at first, one would expect this system to be bistable as well and, as a consequence, to stabilize at the high state until the next interaction occurs. However, at a system level, two phenomena occur simultaneously after the NC valve has been opened: i) pressure $p_\textrm{vent}$ in the bottom chamber decreases in time as air vents to the atmosphere through the resistance $R_\textrm{vent}$ (Fig.~\ref{fig3}I and Fig.~\ref{fig3}J, yellow); ii) pressure in the top chamber oscillates between $\sim\SI{1}{\kilo\pascal}$ and $\sim\SI{35}{\kilo\pascal}$, because of the hysteretic valve oscillating. Given these two phenomena, as long as the difference between the bottom and top chamber is greater than the snap-back pressure of the shell ($\sim\SI{-10}{\kilo\pascal}$), the shell is in the stable snapped branch, because the stability of the shell is determined by the pressure difference between its top and bottom surface (Fig.~\ref{fig2}E). When the pressure difference between the bottom chamber and the top chamber decreases below $\sim\SI{-10}{\kilo\pascal}$, the shell snaps back spontaneously, without external interactions (Fig.~\ref{fig_shortterm_details} and Video~2). As a consequence, the system self-resets to the low state (Fig.~\ref{fig3}J, black).

Note that the snap-back of the shell is not instantaneous, but instead, it lasts approximately \SI{45}{\second}. This is a consequence of the high venting resistance $R_\textrm{vent}$ ($32$ Gauge, \SI{0.25}{\inch} needle) limiting airflow. Once the shell initiates the snap-back, it fights against this high venting resistance, effectively compressing the air in the bottom chamber. The signature of this effect, which counters the snap-back of the shell, can be seen in the sudden increase of the pressure $p_\textrm{vent}$ in the bottom chamber when the shell initiates the snap-back (Fig.~\ref{fig_shortterm_details} and Fig.~\ref{fig3}J, yellow). When the shell is completely snapped back, at around \SI{90}{\second} in the experiment, $p_\textrm{vent}$ finally drops to zero (Fig.~\ref{fig3}J, rest symbol). Nevertheless, we identify the initiation of the snap-back as the instant when the behavior of the system self-resets to the initial state, because this event is associated with the stroke of the actuator returning to approximately the initial value (Fig.~\ref{fig3}J, black). 

Therefore, the short-term memory circuit temporarily stores information by setting the memory element to the snapped state. After a specific memory-retention time, the system spontaneously self-resets to the rest state (as it was before the interaction occurred) and hence the memory of the occurred stimuli fades away. The memory-retention time can be tuned by selecting different values of venting resistance: higher venting resistance results in slower discharge and, therefore, longer memory-retention time. Within the set of venting resistances that we tested, the memory-retention time ranged from \SI{11}{\second} to \SI{71}{\second} (Fig.~\ref{fig_varyshortterm}).

\paragraph*{Fluidic touch sensing via normally open and normally closed kinking tubes}

\begin{figure}[t!]
\includegraphics[width=18cm]{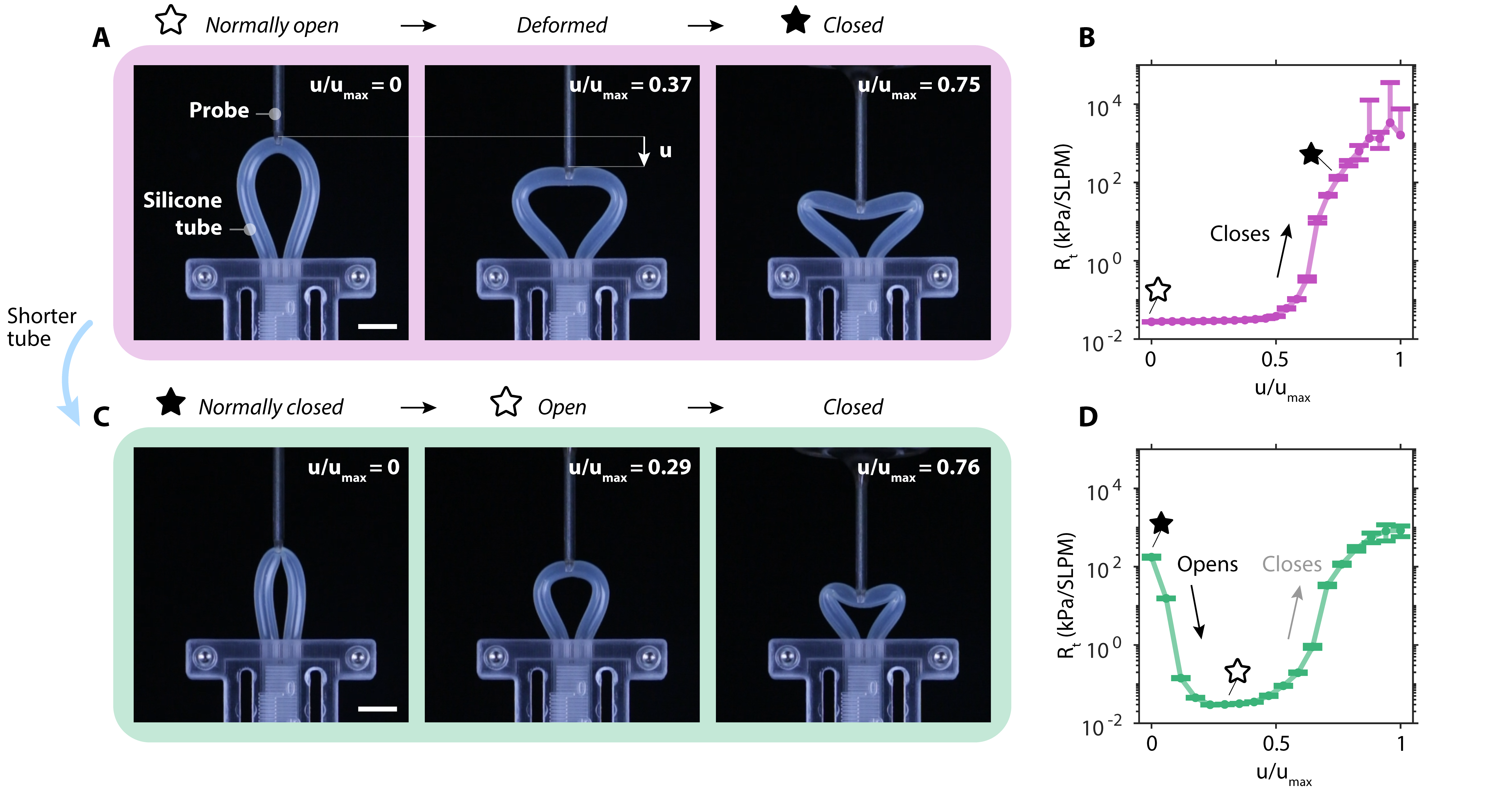}
\par\medskip
    \caption{\label{fig4}\textbf{Fluidic touch sensing via normally open and normally closed kinking tubes.} A silicone tube (inner diameter \SI{3}{\milli\meter}, thickness \SI{1}{\milli\meter}, and length \SI{70}{\milli\meter}) is bent \SI{180}{\degree} and constrained at the inlet and outlet in a rigid holder. (\textbf{A}) A probe compresses the tube by moving vertically with a displacement $u$, first inducing the tube's deformation and then forming two kinks that close the channel. (\textbf{B}) The fluidic resistance of the tube $R_{\textnormal{t}}$ sharply increases when the two kinks form, indicating the transition from the open state to the closed state. (\textbf{C}) A shorter tube of length \SI{50}{\milli\meter} displays a kink in its rest state (therefore, it is normally closed), it unkinks (opens) when compressed, and it forms two kinks when further compressed (it closes again). (\textbf{D}) The fluidic resistance of the normally closed tube as a function of the probe displacement shows an initial drop (when the tube unkinks, opening) with a subsequent increase (when the two kinks form, closing). Data in \textbf{B} and \textbf{D} are reported as mean $\pm$ standard deviation. All scale bars are \SI{1}{\centi\meter}.}
\end{figure}

So far, we could interact with the long- and short-term memory circuits through NO and NC solenoid valves, requiring a \SI{24}{\volt} signal to close and open the valves (Methods). With the goal of developing fully-fluidic machines that can interact with the surroundings, we aim to develop NO and NC touch sensors harnessing the kinking behavior of soft tubes \cite{kink_valves}. We build a NO sensor by bending a commercial elastomeric tube with inner diameter \SI{3}{\milli\meter}, thickness \SI{1}{\milli\meter}, and length \SI{70}{\milli\meter} (Fig.~\ref{fig4}A). We mechanically compress the tube to a maximum displacement $u_\textrm{max}$ corresponding to the top part of the tube touching the bottom rigid holder. At the same time, we measure the fluidic resistance of the tube as the ratio between the inlet-outlet pressure difference and the flow through the tube. While loading, we observe an initial deformation, followed by a sudden formation of two kinks (Fig.~\ref{fig4}A). The flow resistance is relatively low ($\sim\SI{1.5e-2}{\kilo\pascal\per\SLPM}$) when the tube is not deformed ($u/u_\textrm{max}=0$) (Fig.~\ref{fig4}B, white star). The resistance stays at approximately the same value until half the maximum probing displacement ($u=0.5\cdot u_\textrm{max}$). Then, the resistance suddenly increases to $\sim\SI{1e1}{\kilo\pascal\per\SLPM}$ when $u=0.6\cdot u_\textrm{max}$. This increase corresponds with the formation of two kinks that block air flow (Fig.~\ref{fig4}A, black star). The resistance then increases exponentially with increasing probing displacement, as the kinks sharpen further. We call this high-resistance state the closed state (Fig.~\ref{fig4}B, black star), as leakage is limited. So, effectively, the tube is a sensor that transduces information from the mechanical domain (that is, the compressing interaction) into the fluidic domain (that is, the resistance of a fluidic channel).

To build the NC sensor, we found that we can act on a single design parameter of the tube. In fact, by reducing the length of the tube from \SI{70}{\milli\meter} to \SI{50}{\milli\meter}, a kink spontaneously forms when the tube is not probed (Fig.~\ref{fig4}C, black star). This is accompanied by the resistance being relatively high ($\sim\SI{1e2}{\kilo\pascal\per\SLPM}$) (Fig.~\ref{fig4}D, black star): the sensor is in a normally closed state. By compressing, the tube unkinks (Fig.~\ref{fig4}C, white star), opening the channel and suddenly decreasing the resistance to $\sim\SI{1.5e-2}{\kilo\pascal\per\SLPM}$ (Fig.~\ref{fig4}D, white star). So, the shorter tube is an NC sensor that opens upon interaction. In addition, by further probing the tube after the opening event, it closes again, as two kinks form (Fig.~\ref{fig4}C).

Therefore, we obtained a touch sensor that, depending on the design parameters, can be in a NO or a NC setting. Interestingly, the NO and NC mechano-fluidic sensors transduce qualitatively equivalent mechanical inputs (compression in one direction) in distinct fluidic information, consisting of closing and opening of channels.

\paragraph*{Integration of memory and sensing for programmable behaviors upon user interactions}

So far, we introduced a number of separate ingredients that, when integrated together, we will show can result in programmable behaviors. We have \emph{i)} the forward crawling platform (Fig.~\ref{fig1}); \emph{ii)} the mechano-fluidic memory element (Fig.~\ref{fig2}), responsible for providing memory of a stimulus in the form of change in behavior; \emph{iii)} the short- and long-term memory circuit designs (Fig.~\ref{fig3}), that read fluidic stimuli in the form of opening and closing of channels; \emph{iv)} the touch sensors (Fig.~\ref{fig4}), to transduce mechanical stimuli into fluidic ones.

\begin{figure}[t!]
\includegraphics[width=18cm]{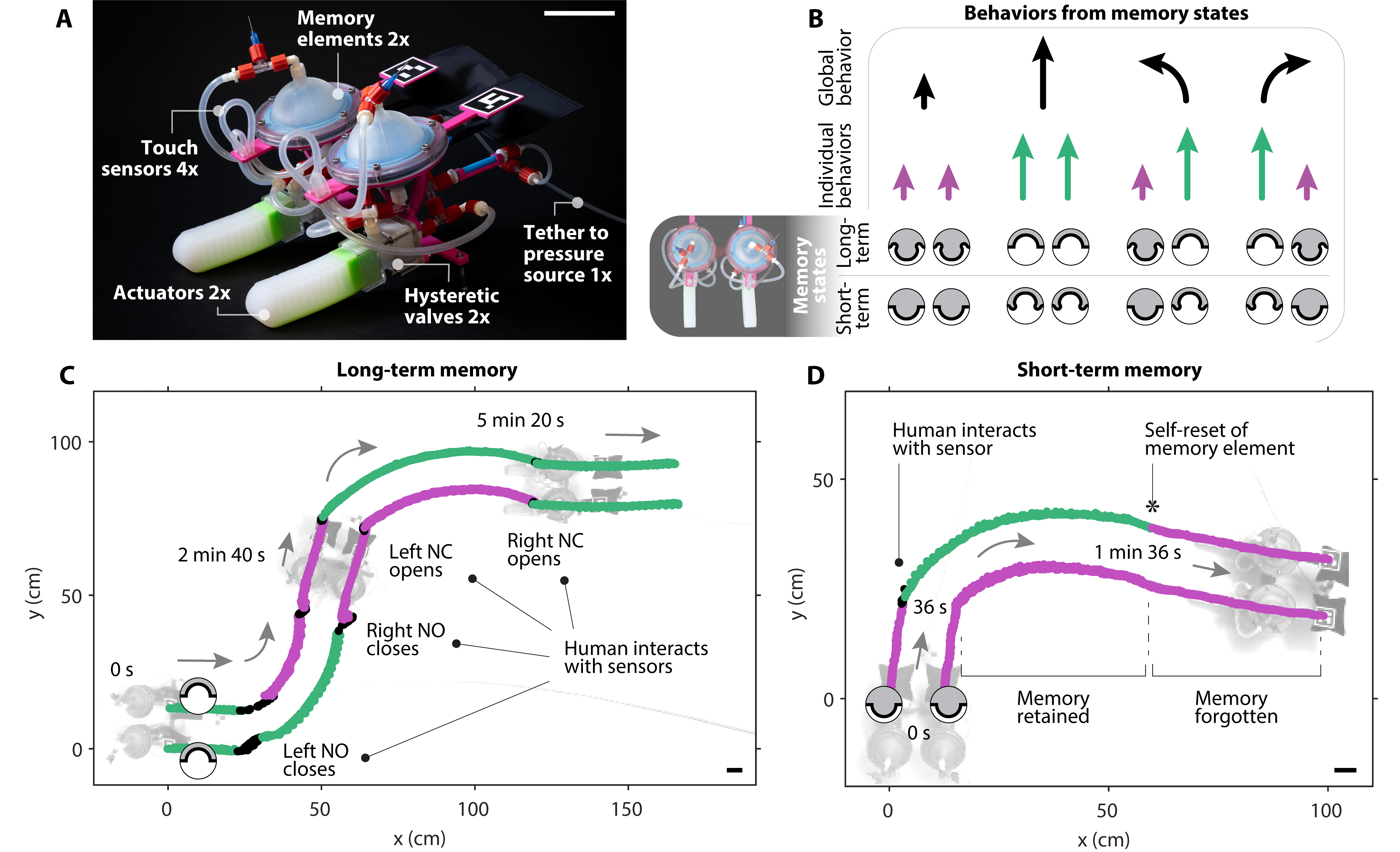}
\par\medskip
    \caption{\label{fig5} \textbf{Integration of memory and sensing for programmable behaviors upon user interactions.} (\textbf{A}) We build a two-limb machine by mirroring our single-limb platform in Figure~\ref{fig2}. The machine is powered by one pressure tether and is equipped with two hysteretic valves, two actuators, two memory units, and four touch sensors (of which two NO and two NC). (\textbf{B}) The individual states of the two memory elements impact the individual behavior of the two halves of the machine, resulting in four different global behaviors of the machine. (\textbf{C}) With long-term memory settings, the machine locomotes in an arena and switches between steering and forward behaviors when a human operator interacts with the touch sensors, by closing the NO and opening the NC sensors. (\textbf{D}) With short-term memory settings, the machine switches to steering behavior when a human interacts with the sensor, and after $\sim\SI{1}{\minute}$, it spontaneously returns to the initial forward-locomoting state. All scale bars are \SI{5}{\centi\meter}. See also Video~S3.}
\end{figure}

Now, we integrate these ingredients in a locomoting soft machine that senses external mechanical cues and reacts accordingly (Fig.~\ref{fig5}A). The machine has two mirrored sides. Each side is built starting from the single-actuator crawling platform, with the memory element and the NO and NC touch sensors placed on top. Internally, the fluidic circuit of each mirrored side is identical to the circuit in Figure~\ref{fig3}. The two sides of the machine are physically connected with a bearing, to allow rotation of the two sides relative to each other, as the crawling behavior relies on the rotation induced by the bending actuator (Fig.~\ref{fig_robot_design}). In total, the machine integrates two bending actuators, two hysteretic valves, two memory elements, four touch sensors, and a single tether connected to a pressure source of \SI{1.3}{\bar}.

With two bistable memory elements, the machine has four stable locomoting behaviors (Fig.~\ref{fig5}B): \emph{i)} forward slow, when both individual capacitors are set to the large volume state; \emph{ii)} forward fast, when both are set to the small volume state; \emph{iii)} steer left, when the right capacitor is set to small volume state and the left to large volume state, as the right side of the machine moves at a faster speed compared to the left side, causing global steering; \emph{iv)} steer right, when the right capacitor is set to large volume and the left to small.

In Figure~\ref{fig5}C, we report the trajectory of the machine locomoting in an arena when receiving touch cues from a human user, with both circuits set to the long-term memory configuration from Figure~\ref{fig3}A. The machine starts with both capacitors set to small volume (both shells set to the rest state). Hence, the machine starts locomoting with the fast-forward behavior. Then, the human interacts with the left NO sensor, closing it. After the interaction, the machine steers to the left because the left memory element is set to the large-volume state, and the resulting speed of the left side is lower than the right side. The memory is retained until a new mechanical cue occurs. After the operator closes the right NO sensor by touching it, the machine displays the slow-forward behavior. With the next two interactions, the machine steers to the right and then goes back to the initial fast-forward behavior (Video~3). Additionally, we can interact with two sensors at the same time, closing one and opening the other, to directly switch between sterring left and steering right (Video~3).

When we set the internal circuit to the short-term memory configuration of Figure~\ref{fig3}G by flipping the orientation of the memory element and increasing the venting resistance (replacing it with a $34$ Gauge, \SI{0.25}{\inch} needle), the machine also displays short-term memory (Fig.~\ref{fig5}D). The machine starts with the slow-forward locomoting behavior. After the user interacts with the left sensor, the left side is set to the high-speed state, and the machine displays the right-steering behavior. After approximately one minute, the left memory element self-resets to the initial state, and the machine returns to the forward-locomoting behavior (Video~3). Therefore, the memory of the interaction is retained for approximately one minute so that the machine steers $\sim\SI{90}{\degree}$. After this amount of time, the memory of the stimulus is forgotten, and the machine goes back to the default behavior, which is moving forward without steering. Short-term memory implies that we can design a preferred behavior, in this case, moving forward, and a temporary behavior, such as steering, that, upon interaction, overrides the default one for a determined amount of time.

\paragraph*{Autonomous obstacle avoidance through short-term memory and insect-inspired sensing}

So far, the machine was designed for only a specific kind of interaction, where a human user touches the sensor. To demonstrate the potential of integrating memory and sensing for autonomous behaviors, where the machine itself senses the environment without external inputs from a human, we design an antenna taking inspiration from insects. Often, insects sense the external world with antennae made of relatively rigid flagella and softer pedicel-flagellum junctions (hinges). The hinges are the crucial site for mechanoreception, as they are equipped with dense mechanosensory structures \cite{insect_antenna_chapter}. When the flagellum bends, the receptors in the hinge sense an increase in bending torque, so that the insect detects obstacles \cite{locust_antenna_object} or changes in the wind direction \cite{locust_antenna_object, insect_antenna_wind}.

\begin{figure}[t!]
\centering
\includegraphics[width=12.5cm]{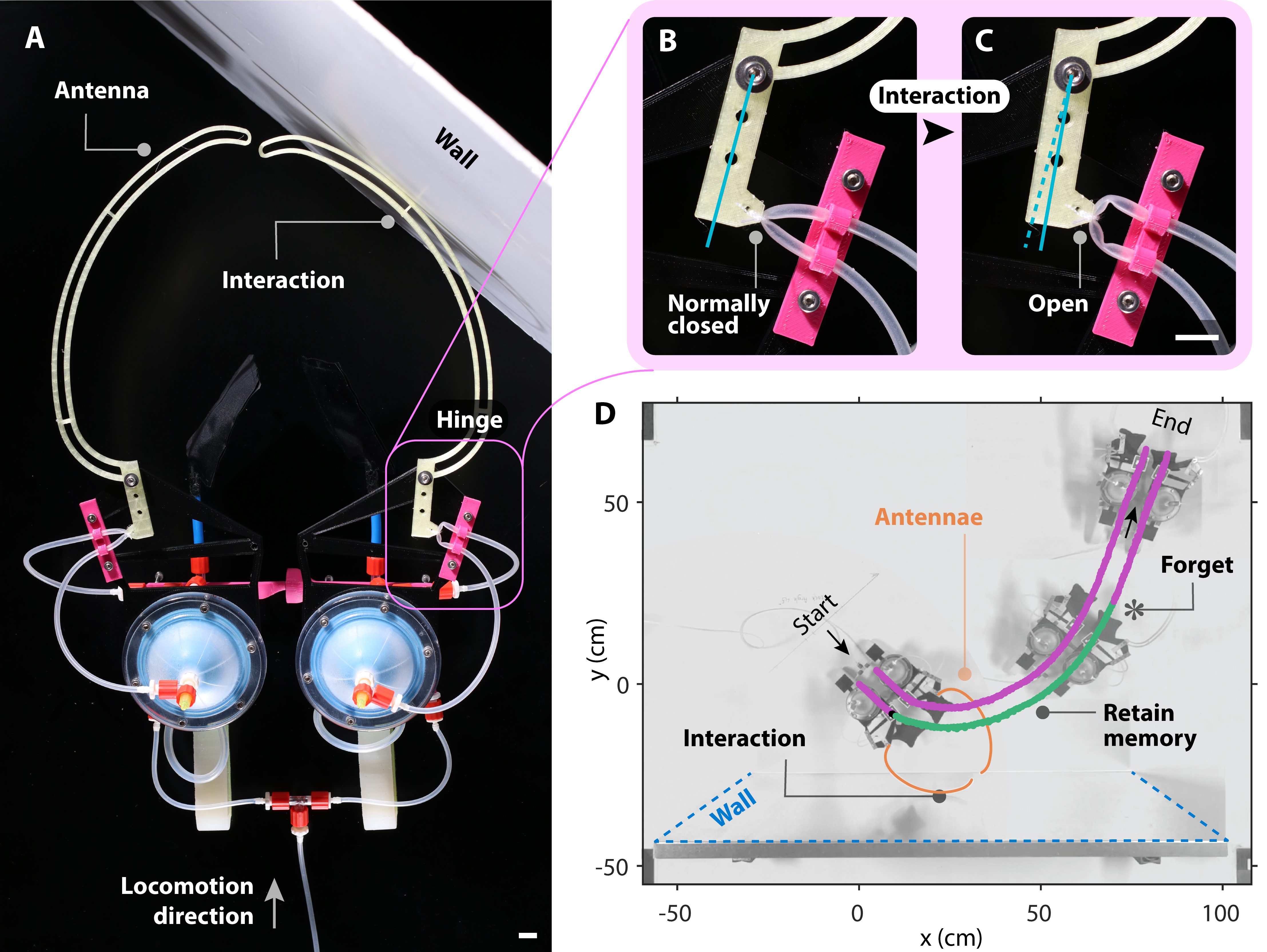}
\par\medskip
    \caption{\label{fig6} \textbf{Autonomous obstacle avoidance through short-term memory and insect-inspired sensing.} (\textbf{A}) We place two 3D-printed antennae on board the machine, allowing rotation around a hinge upon external interaction with an obstacle, such as a wall. (\textbf{B}) At the hinge, the antenna is equipped with a kinked tube as an NC sensor. (\textbf{C}) The tube unkinks upon interaction, effectively sensing the bending moment applied to the antenna and transducing it into fluidic information (drop in fluidic resistance). (\textbf{D}) The autonomous machine senses the presence of a wall, temporarily retains the memory of the interaction by steering away, and finally, forgets the memory of the interaction, going back to the default forward-locomoting behavior. All scale bars are \SI{1}{\centi\meter}. See also Video~S4.}
\end{figure}

We develop an artificial antenna composed of a rigid 3D-printed flagellum that rotates around a hinge (Fig.~\ref{fig_antenna_design}) when it interacts with an external obstacle, such as a wall (Fig.~\ref{fig6}A). Crucially, when the flagellum rotates, a kinked tube placed at the hinge unkinks (Fig.~\ref{fig6}B), acting like the NC touch sensor from Figure~\ref{fig4}C. Hence, we localize at the soft hinge the transduction from external mechanical stimuli into internal fluidic information.

We let a machine equipped with two antennae and two short-term memory circuits walk toward a wall with a \SI{45}{\degree} angle of attack (Fig.~\ref{fig6}D and Video~4). When the right antenna touches the wall, it rotates around the hinge, unkinking the tube (Fig.~\ref{fig6}D, label \apos{Interaction}). This sensing event results in the memory element snapping to the snapped state, following the same mechanism as previously introduced in Figure~\ref{fig3}E,F. Then, the machine steers away from the wall for approximately one minute, because the memory of the interaction is retained (Fig.~\ref{fig6}D, label \apos{Memory retained}). After this memory-retention time, the memory element self-resets to the initial state, in the same fashion as explained in Figure~\ref{fig4}E-H. At this moment in time, the machine forgets the previous interaction with the wall, returning to the forward-locomoting behavior (Fig.~\ref{fig6}D, label \apos{Forget}).

\section*{Conclusions and outlook}

By embodying mechano-fluidic memory in soft machines we obtained programmable behaviors upon interaction with the environment without requiring electronics or software for control. We introduced memory in a soft, self-oscillating crawler through the bistability of a physical parameter of the machine, the capacitance, by leveraging a bistable elastic shell. With long- and short-term memory circuits and kinking tubes as touch sensors, the machine could detect interactions with a user and obstacles, and consequently switch behavior between walking straight and steering.

We instilled memory and feedback effects by coupling together selectively a low number of components (shells and tubes) that display highly nonlinear behavior. This approach parts ways from general-purpose robotics, which instead typically relies on a large number of relatively simple components in a central computer, in the range of billions of nanoscale transistors \cite{moors_law_forever}. We envision that this approach of integrating a limited number of nonlinear components will prove effective for task-specific applications \cite{sitti_physical_intelligence} where specialized tasks are pre-specified and robustness is of the essence, as it reduces the overall design complexity in terms of number of components. For instance, in biomedical applications, microrobots could autonomously navigate inside the human body \cite{sitti_biomedical} and detect cancer tissues by sensing a difference in stiffness \cite{tumor_stiffness, tumor_stiffness_capsule} without carrying microelectronics \cite{microrobots_cohen}.  We also foresee using these mechanical effects as \apos{reflexes} embedded in the structure of machines, to which some tasks can be offloaded from a central controller. For example, in space applications, robots could delegate the autonomous exploration task to their mechanical body, preventing the breakdown of the locomoting apparatus in case of solar storm events \cite{challenges_space}.

Our proposed approach will particularly benefit by carefully tuning the interactions \cite{luuk_autonomy}, both internal to the machines and with the external environment. However, this potential will come with the associated downside of an increased complexity of the design process that takes into account the interactions. While here we specifically focused on memory effects, other complex behaviors are at reach through mechanical interactions, as demonstrated by initial results on self-learning mechanical circuits \cite{prakash_learning} and many-agent cooperation through implicit mechanical couplings \cite{mannus}.  This \apos{integrative mechanics} approach is the artificial parallel to integrative biology, where complex dynamic behaviors emerge from the interplay between interactions within the agent and with the environment in which the agent is situated. These specialized, distributed, and redundant interactions found in biological systems enable high levels of robustness, while not sacrificing functionality: autonomous machines have the potential to tap into this vast complexity as well.

\newpage

\section*{Methods}

\subsection*{Design and manufacturing}
\vspace{8pt}

\subsubsection*{Hysteretic valves, bending actuators, soft resistors, and inextensible capacitors}

The design and the manufacturing procedure of the hysteretic valves are based on the original design we previously introduced \cite{luuk}. We scaled up the original design by a factor of $2$ in all dimensions, and we introduced a notch in the rigid holders in the same way as described in detail in our previous work \cite{coexistence}.

The bending PneuNet actuator was previously reproduced by our lab, and both its design and the manufacturing steps are described in detail in our previous work \cite{shibo, pneumatic_coding}. The actuator presents two sides, manufactured via two consecutive injection molding steps. The inflating side with hollow chambers is made of DragonSkin 10 (DS10) silicone (Smooth-On). The stiffer side is made of Elite Double 32 (ED32) silicone (Zhermack), with an embedded inextensible grid fabric (Penelope 70/10, Garenenzo). Before injecting ED32, it is essential to wait for DS10 to be partially cured (\SI{3.5}{\hour} of the total \SI{5}{\hour} curing time), to allow for stronger bonding between the two silicones.

The soft resistors, implemented as pre-resistances in all the circuits, are elastomeric hollow cylinders made of Smooth‑Sil 950 (Smooth-On), with outer diameter \SI{5}{\milli\meter}, inner diameter \SI{0.35}{\milli\meter}, and length \SI{40}{\milli\meter}. The soft resistors are manufactured via injection molding, with a two-part outer mold and a metal rod with diameter \SI{0.35}{\milli\meter} as the inner mold.

The inextensible capacitors are made of two TPU-coated nylon sheets (\apos{nylon 70den TPU-coated one side 170g/sqm heat-sealable}, extremtextil) that we heat-seal along a specified path. The path is a rectangle, with an opening on one of the short sides (Fig.~\ref{fig_pouches}). Given the width $w$ and the height $h$, we calculate the geometric volume $V$ of the pouch when inflated as the solution to the paper bag problem \cite{paperbag}:
\begin{equation}
    V = w^3\cdot\left(\frac{h}{\pi w}-0.142\cdot(1-10^{-h/w})\right).
    \label{eq:paperbag}
\end{equation}
The two sheets are heat-sealed using the 3D printer Felix Tec 4. The printer presents a custom-made hot head, previously used by our group \cite{arfaee_pouch}, that consists of a spherical hot nozzle and a spring that ensures even sealing lines. A silicone mat is placed between the printer bed and the sheets, and oven paper is placed on top of the sheets. To the printer, we send a G-code based on an Adobe Illustrator file containing the desired tool path. After sealing, we cut the residual material surrounding the sealing lines and,  in the open side of the pouch, we heat-seal the sheets to a TPU Festo\textsuperscript{TM} tube with a soldering iron at \SI{300}{\degree}. Finally, we place Luer\textsuperscript{TM} connectors (MLRL007-1 Male Luer to 500 Series Barb 3/32" 2.4 mm with Lock Ring FSLLR-3) in the Festo\textsuperscript{TM} tubes.

Throughout the article, we place a fixed pre-capacitor with volume $V_\textrm{pre}=\SI{60}{\milli\liter}$ (with $w=h=\SI{68}{\milli\meter}$) on the machines (except in Figure~\ref{fig_varyfluidicparameters}A-C, where we purposefully vary the pre-capacitance). 

\subsubsection*{Bistable shells}

The bistable shells are defined by four parameters, as illustrated in Figure~\ref{fig_dome_design}A: the thickness $t$, the angle $\alpha$ from the vertical axis, the base width $w$, and the boundary radius $R_\textrm{b}$. Throughout our study, the shells have  $t = \SI{3}{\milli\meter}$, $\alpha = \SI{80}{\degree}$, $w = \SI{55}{\milli\meter}$, and $R_\textrm{b} = \SI{2.28}{\milli\meter}$. In addition, the shells have an outer notch and rim that allows for centering when being mounted in the rigid holders. The rigid holder clamps the shell along the outer rim, held together with screws (Fig.~\ref{fig_dome_design}B). The rigid holders compress the silicone rim of \SI{0.95}{\milli\meter} ($\sim30\%$ of its thickness).

The shells are manufactured via injection molding of Smooth‑Sil 950 silicone (Smooth-On) using a two-part outer mold. The outer molds were printed in VeroClear (Stratasys) with a PolyJet 3D printer (Eden260VS, Stratasys). Before molding, we sprayed a thin layer of release agent (Ease release 200, Smooth-On) on the inner surface of the molds to ease demolding after curing.

\subsubsection*{Crawling soft machines}

The single-actuator crawling soft machine reported in Figure~\ref{fig1} consists of various individual components (Fig.~\ref{fig_robot_design}). The first component consists of two rigid 3D-printed parts that hold the hysteretic valve in place. The part on the outlet side of the valve has a cavity where the bending actuator is press-fit. Both inlet and outlet parts have a hole with diameter \SI{5.5}{\milli\meter} that, after tapping with a 1/4-28 UNF tap, allows for the insertion of a threaded Luer\textsuperscript{TM} connector (Luer\textsuperscript{TM} quick-turn tube coupling 1/4"-28 UNF). These Luer\textsuperscript{TM} connectors allow for the other modules (heat-sealed pouches, resistors, and shells) to be quickly connected. In addition, on the inlet side, we attach a 3D-printed adapter, on which we place screws that act as rigid legs of height \SI{25}{\milli\meter}. 

The two-actuator crawling soft machine (Fig.~\ref{fig5}) consists of two identical copies of the single-actuator crawler. The only difference lies in the adapter for the rigid legs, which is designed to accommodate a bearing (Fig.~\ref{fig_robot_design}). The bearing allows for the rotation of the two single-actuator sides with respect to each other, decoupling the two rotational degrees of freedom. In this way, when the two actuators activate at different frequencies and amplitudes, the two sides locomote at different speeds. Without the bearing, when one side actuates, the other side would drag along.

\subsection*{Experimental setups}
\vspace{8pt}

\subsubsection*{Measuring fluidic quantities}

To regulate and measure the fluidic quantities of interest (pressure and flow), we use a custom-made acquisition setup previously used by our group \cite{luuk, coexistence, pneumatic_coding}. The setup is based on the National Instruments IN USB-6212 input/output board, with software developed in-house. A proportional pressure regulator (Festo\textsuperscript{TM} VEAB-L-26-D18-Q4-V1-1R1) is controlled using an analog port of the board. The regulator is connected to an upstream precision pressure regulator (Festo\textsuperscript{TM} LRP-1/4-10), connected to the building pressure source. The setup has various analog input ports, which read the voltage from the pressure sensors (NXP MPX4250DP) and flow sensors (Honeywell AWM5101VN, Honeywell AWM5104VN). The fluidic connection between the parts (regulators, valves, sensors, and samples) is implemented with silicone tubes (Rubbermagazijn 2x4mm and 3x6mm), Festo\textsuperscript{TM} tubes (PUN-6X1-BL), and Luer\textsuperscript{TM} connectors (male-female, Luer-to-barb, and T connectors).

In the benchtop fluidic experiments in Figure~\ref{fig1}H,I, Figure~\ref{fig3}, Figure~\ref{fig_varyfluidicparameters}A,B,D,E, and Figure~\ref{fig_varyshortterm}, the soft machine is held in place with the actuator not interacting with the ground, free to bend mid-air. Detecting an ArUco marker placed at the tip of the actuator allows for determining the vertical stroke of the actuator when it does not interact with the ground. The normally closed and normally open \SI{24}{\volt} solenoid valves (SMC Solenoid Valve VDW250-5G-2-01F-Q) used in the benchtop experiments are controlled through digital output ports of the input/output board.

\subsubsection*{Compressing the fluidic touch sensors}

To measure the fluidic resistance of the kinking tubes as a function of their deformation (Fig.~\ref{fig4}), we make use of a fluidic analog of a resistive voltage divider circuit. Using the electronic-fluidic analogy \cite{luuk}, we treat pressure as voltage. The kinking tube is placed in series with a resistor (a needle) that vents to atmosphere. This resistor has a fixed known resistance $R_0\approx\SI{0.5}{\kilo\pascal\per\SLPM}$ that we measured separately as the slope of the characteristic pressure-flow curve of the needle. We then connect the kinking tube to a pressure regulator upstream (\SI{1.5}{\bar}), and we measure the pressure before the kinking tube $p_{\textrm{in}}$ and the pressure after the kinking tube $p_{\textrm{out}}$ from which we can derive the resistance of the kinking tube $R_\textrm{t}$:
\begin{equation}
   R_\textrm{t} = R_0 \cdot \left( \frac{p_{\textrm{in}}}{p_{\textrm{out}}} - 1 \right).
    \label{eq:paperbag}
\end{equation}
The pressure data is acquired for various static deformation states of the tube. The tube is deformed in increments of \SI{1}{\milli\meter} using a rigid probe, controlled with a tensile-testing machine Instron 5965. The starting moment of each Instron compression is triggered by a digital output signal from the fluidic setup, ensuring that the acquisitions in both setups are synchronized. In Figure~\ref{fig4}, each datapoint is the average fluidic resistance (with standard deviation) over \SI{5}{\second} of this static compression condition.

\subsubsection*{Obtaining the pressure-volume curves}

To obtain the pressure-volume curves of the capacitors in Figure~\ref{fig2} and Figure~\ref{fig_pVcurves}, we quasi-statically inject a controlled amount of water into the capacitors. We mechanically connect a syringe to the tensile-testing machine Instron 5965, and we fluidically connect it to the capacitors making sure all air bubbles are removed. We ramp the displacement of the tensile-testing machine at a rate of \SI{100}{\milli\meter\per\minute}. Since water is incompressible, controlling the displacement of the syringe means controlling the geometric volume provided to the capacitors. Knowing the diameter of the syringe \SI{26.4}{\milli\meter}, volume is therefore ramped at a rate of \SI{54.74}{\milli\liter\per\minute}. While controlling volume, we measure pressure inside the capacitors with a water-compatible pressure sensor (Honeywell 6DF5G). Since the injected water has substantial mass, to remove the effect of gravity on the pressure-volume curves, we submerge the capacitors in a water tank while performing the tests.

\subsubsection*{Tracking machines' location}

To extrapolate the location of the locomoting machines, we place ArUco markers on board the machines, and we use the Python library OpenCV to detect the ArUcos for each frame of the GoPro videos. We first convert each frame to greyscale using the command \texttt{cv2.cvtColor}. We then correct for warp and distortion of the GoPro camera, using four static ArUco markers placed as a reference at the corners of the canvas, with the command \texttt{cv2.warpPerspective}. Then, we detect the ArUcos with the command \texttt{cv2.aruco.detectMarkers}.

\section*{Resource availability}

\subsection*{Lead contact}
Requests for further information and resources should be directed to and will be fulfilled by the lead contact, Johannes T.B. Overvelde (B.Overvelde@amolf.nl).

\section*{Acknowledgments}

We thank all members of our Soft Robotic Matter Group for the invaluable discussions. We thank Niels Commandeur for technical support. \textbf{Funding:} A.C., T.M., and J.T.B.O. acknowledge the European Union’s 2020 ERC-STG under grant agreement No. 948132. This work is part of the Dutch Research Council (NWO) and was performed at the research institute AMOLF.

\section*{Author contributions}

A.C. and J.T.B.O. proposed and developed the research idea; A.C. and T.M. designed and fabricated the devices; A.C. and T.M. performed the experiments; A.C. performed the data analysis; A.C. made the figures and movies; A.C. and J.T.B.O. wrote the manuscript; J.T.B.O. supervised the research.

\section*{Declaration of interests}

The authors declare no competing interests.

\section*{Supplemental information}

Document S1. Figures S1–S9\\
\noindent{Video S1. Memory through a bistable shell, related to Figure 2}\\
\noindent{Video S2. Long- and short-term memory, related to Figure 3}\\
\noindent{Video S3. Switching behaviors via interactions, related to Figure 5}\\
\noindent{Video S4. Autonomous behavior through antennae, related to Figure 6}

%\temp{References cited in the SI and ONLY in the SI should appear ONLY in the SI, not in the main text.}

\bibliographystyle{Science}
\bibliography{scibib}

\newpage

\setcounter{figure}{0}
\renewcommand{\thefigure}{S\arabic{figure}}
\pagenumbering{arabic}

\section*{Supplemental information}

\begin{figure}[h!]
\centering
\includegraphics[width=12.5cm]{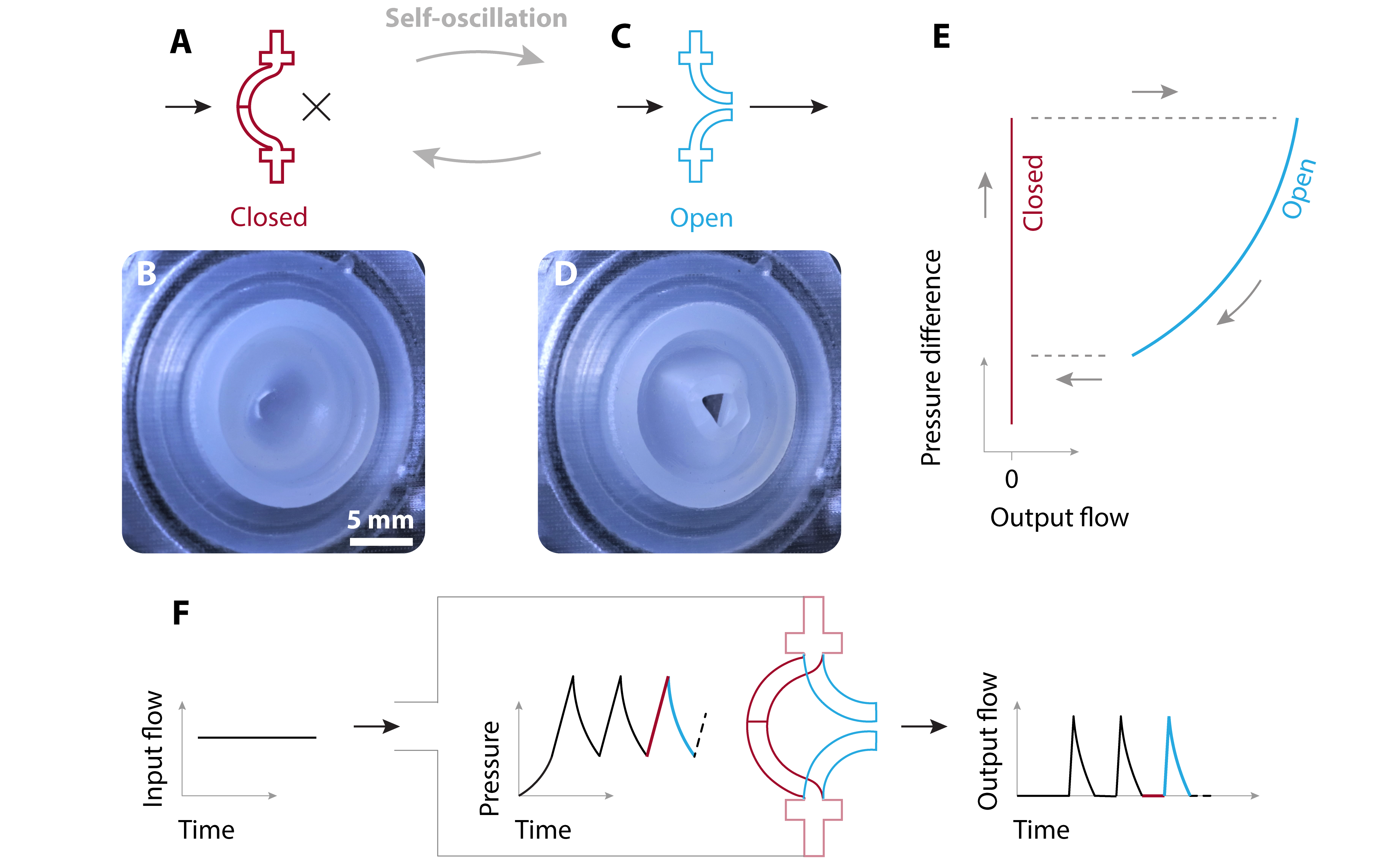}
\par\medskip
    \caption{\label{fig_hysteretic_valve} \textbf{The hysteretic valve self-oscillates given constant input flow.} The hysteretic valve is an elastomeric shell with a slit at the apex. (\textbf{A}) Valve in closed state: when pressure builds up, the slit is closed, and output flow is not provided. (\textbf{B}) Photograph of the closed valve as seen from the outlet side. (\textbf{C}) Valve in open state: when the shell snaps to the collapsed state, the slit opens, allowing flow in output. (\textbf{D}) Photograph of the open valve. (\textbf{E}) Relationship between the output flow and the pressure difference across the valve. When the shell is in the rest state (valve closed), pressure builds up with zero flow in output. When the pressure reaches the snap-through value, the shell snaps (valve opens). When the valve is in the open state, it acts as a nonlinear fluidic resistance that vents flow in output, hence, pressure decreases. When pressure decreases past the snap-back value, the shell snaps back to the rest state (valve closes). (\textbf{F}) When placed in a fluidic circuit with constant flow in input, the valve behaves as a relaxation oscillator \cite{luuk}, continuously transitioning between the closed state and the open state. The upstream pressure oscillates between a low and a high value, and the output flow alternates between zero and positive bursts.  Adapted from Van Laake and Comoretto et al. \cite{coexistence}.}
\end{figure}

\begin{figure}[t!]
\centering
\includegraphics[width=12.5cm]{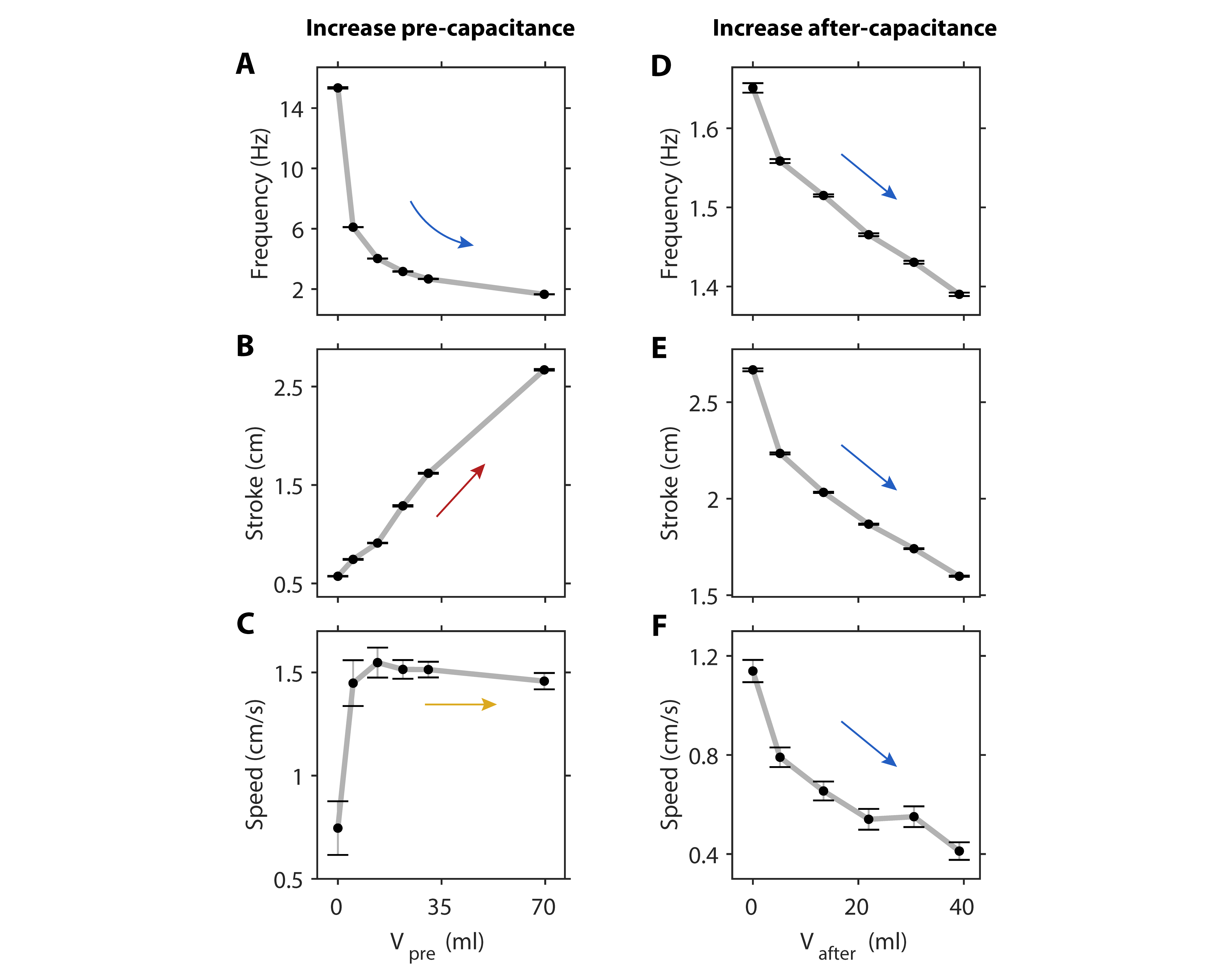}
\par\medskip
    \caption{\label{fig_varyfluidicparameters} \textbf{Response of the crawler in Figure~\ref{fig1} as a function of pre- and after-capacitances.} We vary the pre- and after-capacitances of the circuit in the single-actuator machine (Fig.~\ref{fig1}) by manufacturing various inextensible pouches (Methods) with different height and width (Fig.~\ref{fig_pouches}). In particular, all the pouches have a fixed width of $\SI{30}{\milli\meter}$, and we vary their height from $\SI{30}{\milli\meter}$ to $\SI{150}{\milli\meter}$ in increments of $\SI{30}{\milli\meter}$. In the case of the pre-capacitance, we also test a larger value of volume (\SI{70}{\milli\liter}), as the sum of a pouch with height $\SI{120}{\milli\meter}$ and one with height $\SI{150}{\milli\meter}$. The values of geometric volume reported in this figure result from applying Eq.~\eqref{eq:paperbag}. The frequency and vertical stroke of the actuator are measured in a benchtop experiment, with the actuator not interacting with the surroundings. The speed of the machine is measured by recording the crawler from above, and detecting the ArUco marker (Methods). (\textbf{A}) Frequency of activation of the actuator,  (\textbf{B}) vertical stroke of the actuator, and (\textbf{C}) speed of the machine as a function of pre-capacitance. (\textbf{D}) Frequency of activation of the actuator,  (\textbf{E}) vertical stroke of the actuator, and (\textbf{F}) speed of the machine as a function of after-capacitance. Increasing the pre-capacitance volume causes a decrease in frequency (blue arrow in A) and an increase in stroke (red arrow in B), resulting in a negligible change in speed (yellow arrow in C) for positive pre-capacitance volumes. In contrast, increasing the after-capacitance causes a decrease in both the frequency and stroke of the actuator, leading to a net decrease in the resulting speed of the machine (blue arrows in D, E, F).}
\end{figure}

\begin{figure}[h!]
\centering
\includegraphics[width=10.5cm]{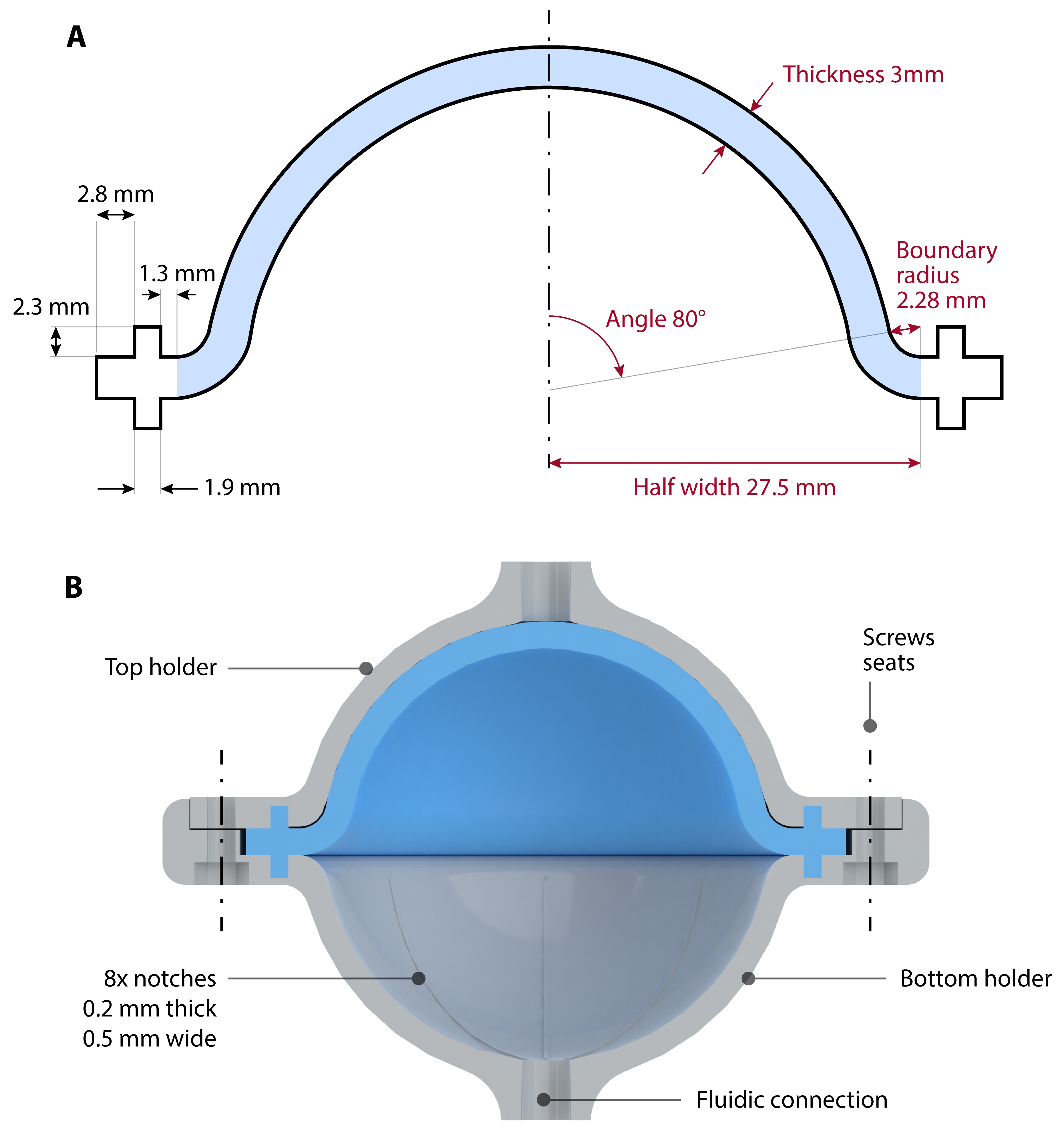}
\par\medskip
    \caption{\label{fig_dome_design} \textbf{Design of the bistable elastic shells.} (\textbf{A}) Schematic of the cross-section of the shell, with the unclamped section highlighted in blue, and the clamped section highlighted in white. The design parameters that affect the behavior of the shell when pressurized are the thickness, the shallowness angle, the width, and the boundary radius (red annotations). The shells used throughout the article have thickness \SI{3}{\milli\meter}, angle \SI{80}{\degree}, width \SI{55}{\milli\meter}, and boundary radius \SI{2.28}{\milli\meter}. Practically, to clamp the shell in a rigid holder, we designed an extension on the outer edge consisting of a rim and a notch (black annotations). (\textbf{B}) Render of a cross-section view of the shell in its holder. The holder presents $8$ notches on the inside of each rigid shell, to ensure that the air provided through the fluidic connections distributes pressure evenly on the surface of the silicone shell. The holder has $6$ holes, placed radially outside the shell, used to clamp the shell in place using M3 screws and nuts.}
\end{figure}

\begin{figure}[h!]
\centering
\includegraphics[width=16.5cm]{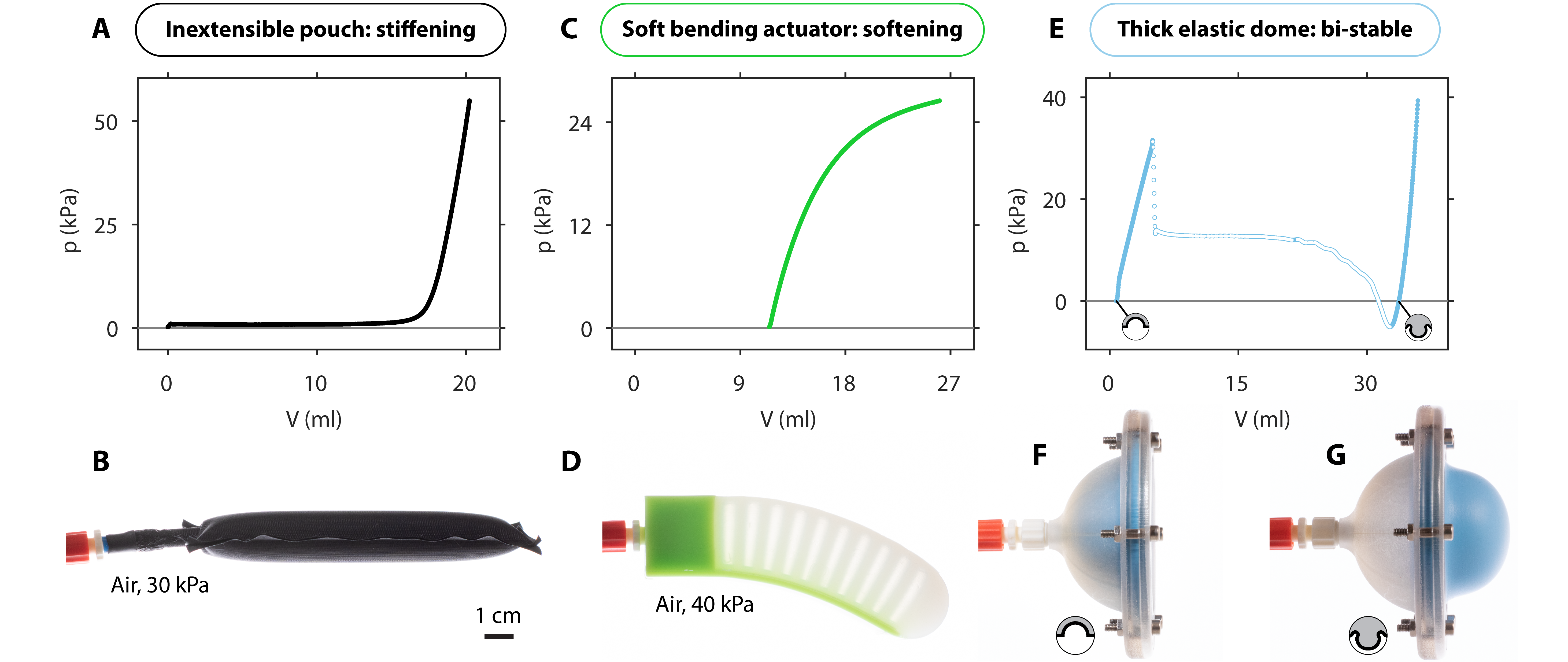}
\par\medskip
    \caption{\label{fig_pVcurves} \textbf{Pressure-volume curves of a pouch, an actuator, and a bistable dome.} We inflate three distinct fluidic capacitors under quasi-static volume-control conditions. We do so by slowly injecting a controlled volume of water that corresponds to the geometric volume inside the capacitors, since water is incompressible; while increasing volume, we measure pressure (Methods). (\textbf{A}) The pressure-volume curve of an inextensible pouch with width \SI{30}{\milli\meter} and height \SI{120}{\milli\meter}. Note that pressure starts to substantially increase only when the volume reaches $\sim\SI{15}{\milli\liter}$. (\textbf{B}) Photograph of the pouch pressurized with air at \SI{30}{\kilo\pascal}. (\textbf{C}) The pressure-volume curve of a soft bending actuator. There is an initial volume at atmospheric pressure corresponding to the geometric volume of the inner chambers. (\textbf{D}) Photograph of the soft bending actuator pressurized with air at \SI{40}{\kilo\pascal}. (\textbf{E}) The pressure-volume curve of our bistable elastic shell. White markers indicate the negative stiffness branch (where an increase in volume causes a decrease in pressure). For increasing volume, pressure is non-monotonous: pressure first increases then decreases, then increases again. Note that the positive-stiffness branches intersect the zero-pressure vertical line at two points: the system is bistable, as it is stable in both these two states without applying pressure. (\textbf{F}) Photograph of the bistable shell in the rest state (volume \SI{0.8}{\milli\liter}, zero pressure).  (\textbf{G}) Photograph of the bistable shell in the snapped state (volume \SI{34}{\milli\liter}, zero pressure). }
\end{figure}

\begin{figure}[h!]
\centering
\includegraphics[width=16.5cm]{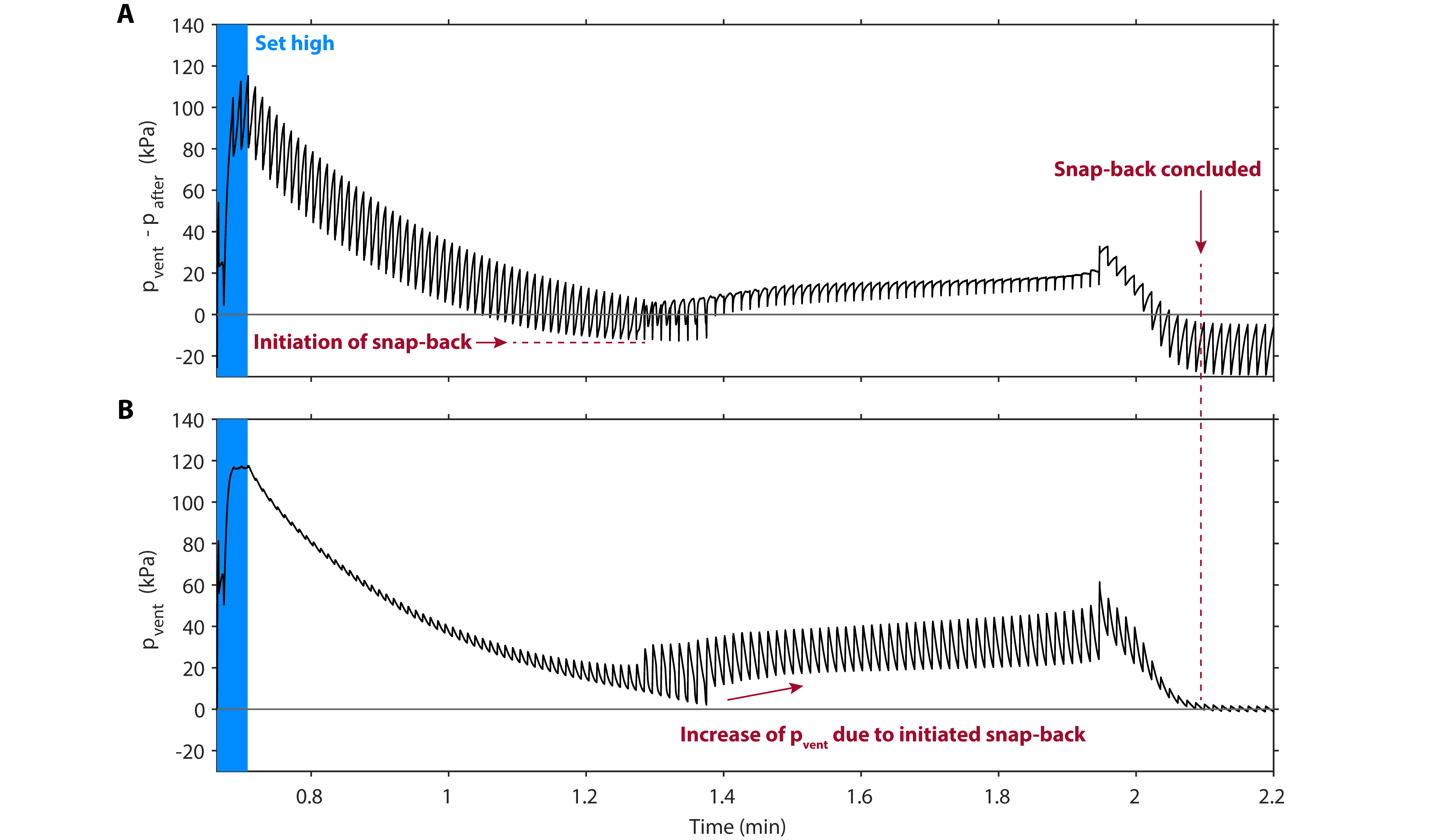}
\par\medskip
    \caption{\label{fig_shortterm_details} \textbf{In the short-term memory circuit in Figure~\ref{fig3}, the spontaneous snap-back is not instantaneous.} (\textbf{A}) The pressure difference between the venting chamber and the chamber after the hysteretic valve ($p_\textrm{vent}-p_\textrm{after}$) decreases in time after the system is set to the snapped state, because air in the venting chamber vents to atmosphere through the venting resistance $R_\textrm{vent}$. Since the hysteretic valve oscillates, this pressure difference oscillates as well, while decreasing. When the pressure difference reaches the snap-back pressure of the shell, the shell initiates the snap-back. (\textbf{B}) After the shell initiates the snap-back, pressure in the venting chamber $p_\textrm{vent}$ increases, because of the relatively high venting resistance $R_\textrm{vent}$ connected to the venting chamber. This happens because the shell deforms towards the venting chamber, effectively compressing the air that does not immediately vent through the high resistance. Approximately \SI{50}{\second} after the initiation of the snap-back, the shell completes the snap-back ($p_\textrm{vent}$ drops to zero).}
\end{figure}

\begin{figure}[h!]
\centering
\includegraphics[width=12.5cm]{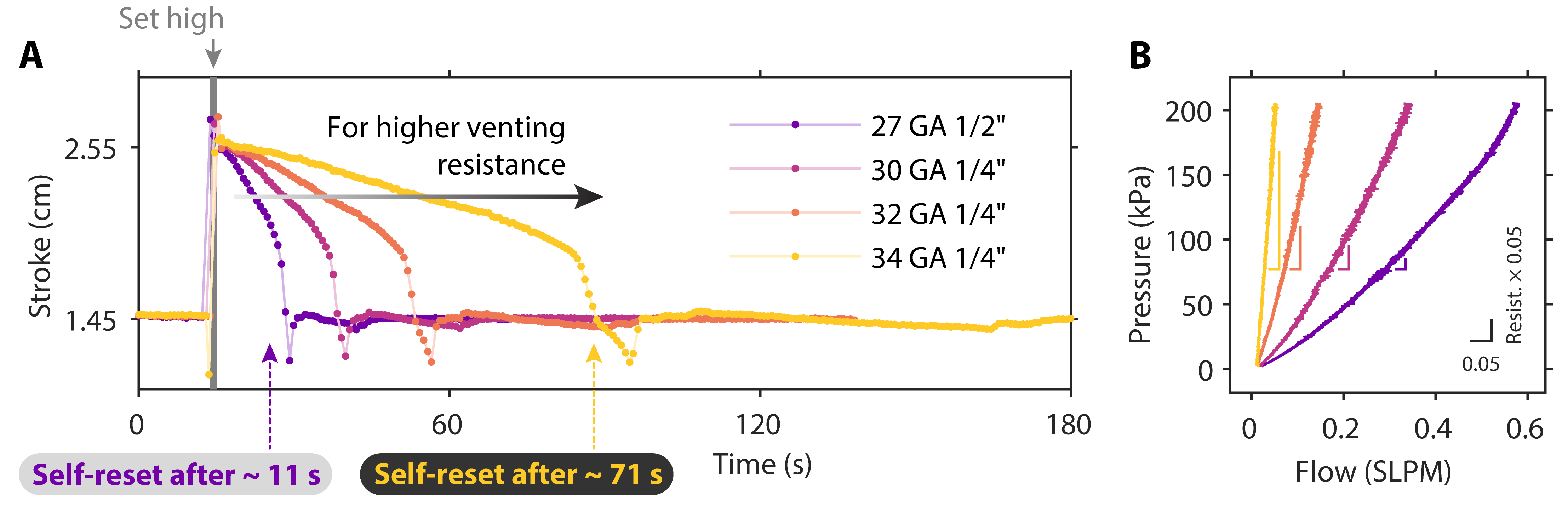}
\par\medskip
    \caption{\label{fig_varyshortterm} \textbf{In the short-term memory circuit in Figure~\ref{fig3}, increasing the venting resistance leads to a higher memory-retention time.} We test the short-term memory circuit (Fig.~\ref{fig3}) for different values of venting resistance (different commercial needles, Metcal). In a benchtop setup (Methods) we measure the vertical stroke of the actuator in time. (\textbf{A}) After opening the normally closed valve, the memory element snaps, and the system is set to the high-stroke state. In the case of the lowest resistance tested ($27$ Gauge, \SI{0.5}{\inch} needle), the system self-resets to the low state (the shell snaps back) after $\sim\SI{11}{\second}$ (purple line). For higher venting resistance, the time required to self-reset (memory-retention time) increases. In the case of the highest resistance tested ($34$ Gauge, \SI{0.25}{\inch} needle), the memory-retention time is $\sim\SI{71}{\second}$ (yellow line). (\textbf{B}) The pressure-flow curves of the tested resistances show that, as expected, the resistance increases with increasing Gauge number of the needles.}
\end{figure}

\begin{figure}[h!]
\centering
\includegraphics[width=11.5cm]{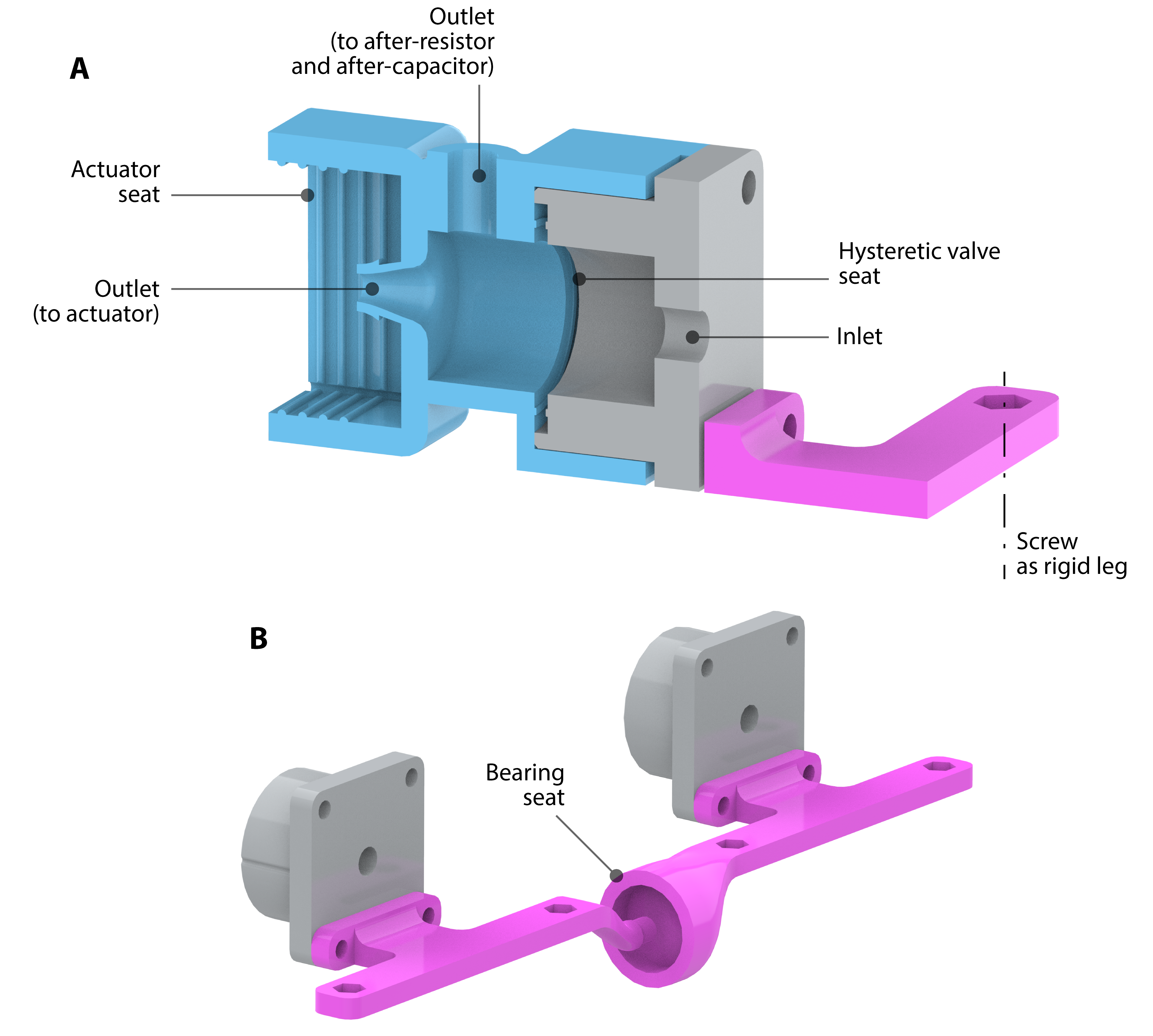}
\par\medskip
    \caption{\label{fig_robot_design} \textbf{Design of the machines.} (\textbf{A}) Cross-section view of the main module to which the rest of the components are connected to assemble the single-actuator machine. The hysteretic valve is seated between the inlet part (grey) and the outlet part (blue). The bending actuator connects to the outlet part via press-fit. An additional part (pink) connects to the inlet part, to allow for screws to be placed as rigid legs. (\textbf{B}) The two-actuator machine presents two mirrored copies of the single-actuator assembly, with the only difference being the rigid-legs parts (pink). These parts accommodate a bearing, to allow relative rotation of the two parts with respect to each other.}
\end{figure}

\begin{figure}[h!]
\centering
\includegraphics[width=14.5cm]{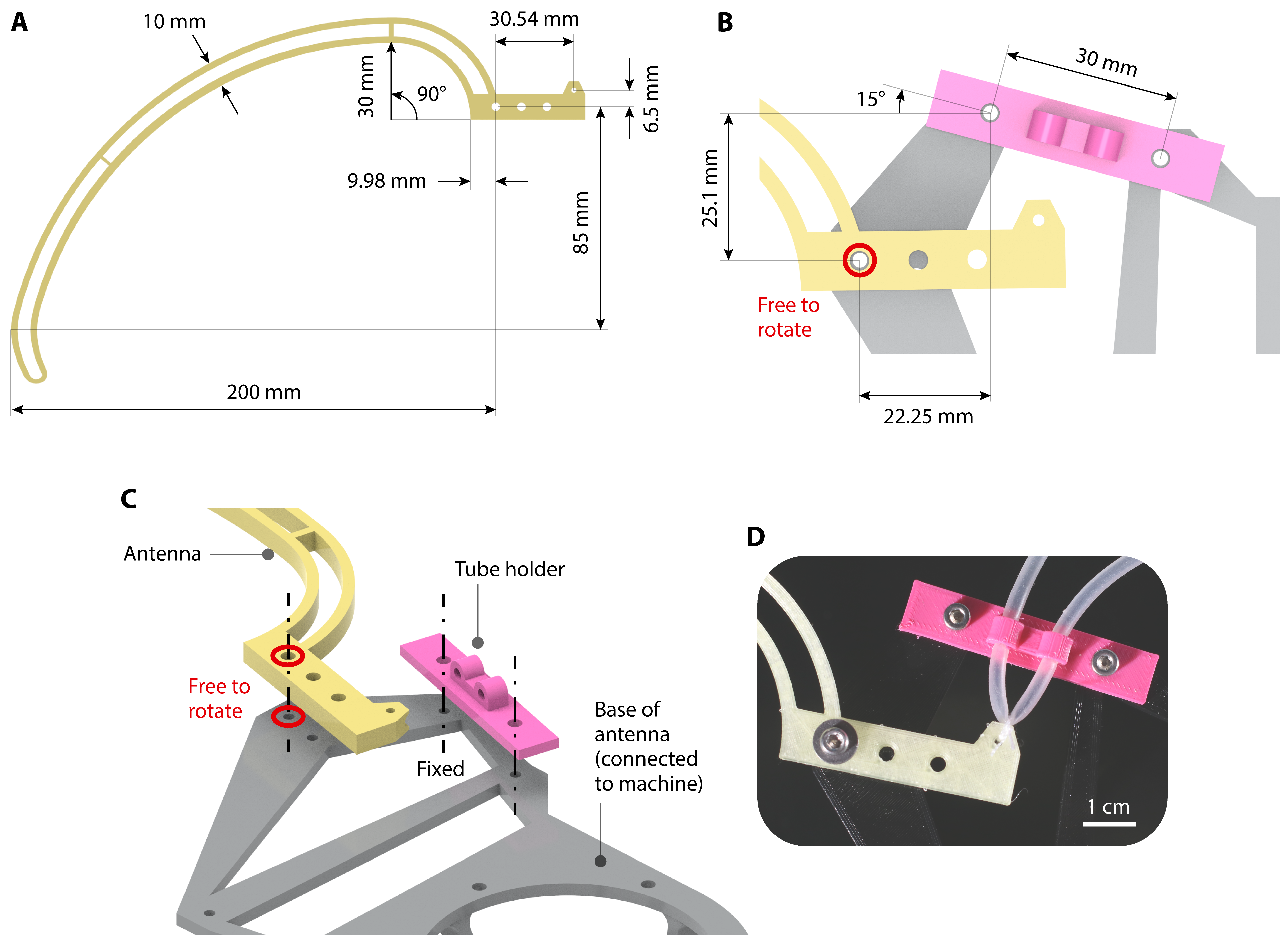}
\par\medskip
    \caption{\label{fig_antenna_design} \textbf{Design of the antenna.} (\textbf{A}) Relevant design parameters of the rigid antenna in Figure~\ref{fig6}. (\textbf{B}) Design parameters of the hinge, with the antenna (yellow) connected to the base (grey) through a loose screw that allows rotation (red mark). (\textbf{C}) Indication of the mounting holes of the antenna (yellow) and the tube holder (pink) to the base (grey). (\textbf{D}) Photograph of the assembled hinge, with the antenna (yellow) firmly attached to the silicone tube (inner diameter \SI{2.5}{\milli\meter}, thickness \SI{0.4}{\milli\meter}, length \SI{30}{\milli\meter}, and inlet-outlet distance \SI{10}{\milli\meter}) with a thin thread.}
\end{figure}

\begin{figure}[h!]
\centering
\includegraphics[width=10.5cm]{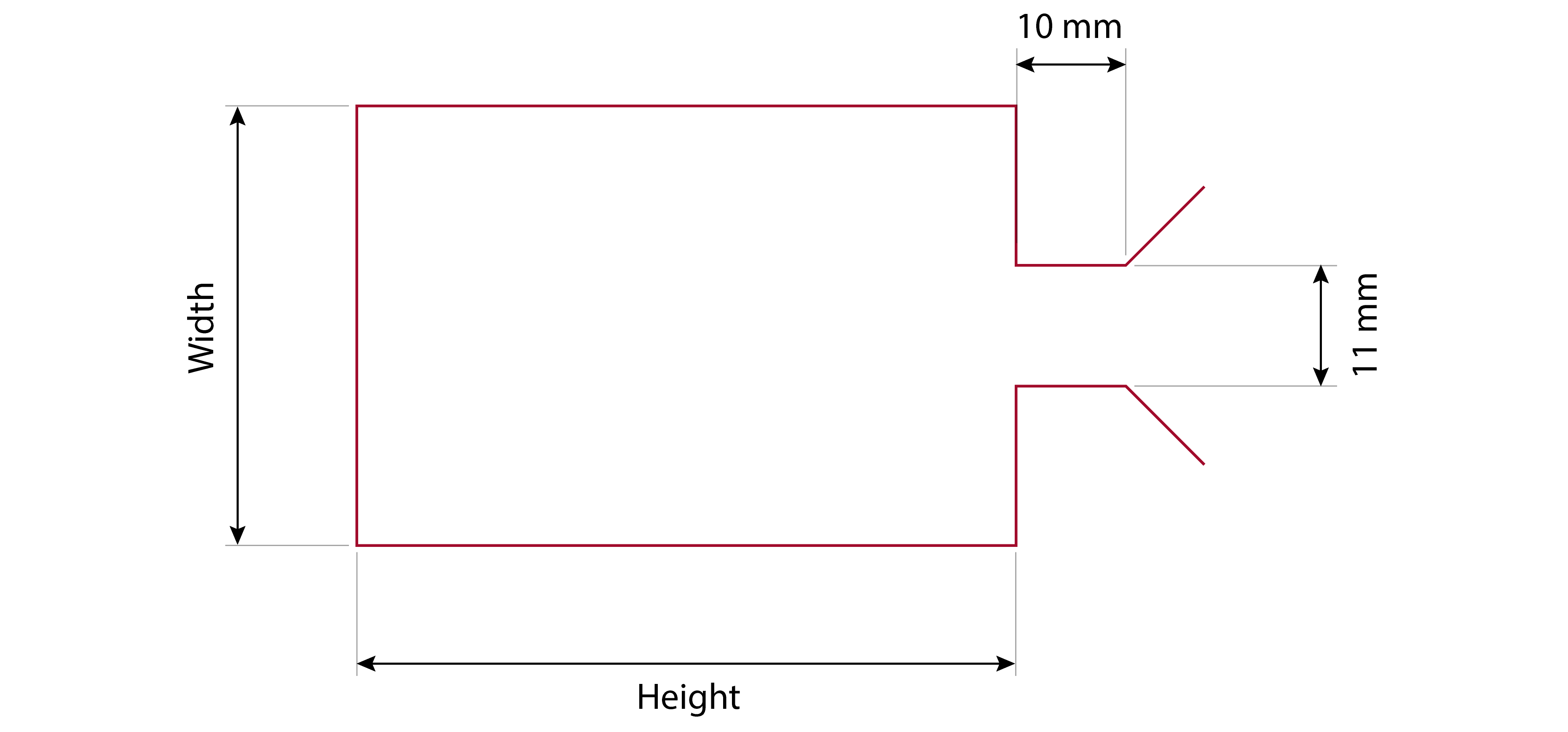}
\par\medskip
    \caption{\label{fig_pouches} \textbf{Design of the inextensible pouches.} The pouches are rectangular, with an opening at one of the shorter sides. An Adobe Illustrator file, containing the red lines reported here, is used to generate the G-code. The 3D printer follows this path to seal two TPU sheets together (Methods).}
\end{figure}

\end{document}